\documentclass[12pt]{article}
\usepackage{epsfig}
\usepackage{amsfonts}
\begin{document}
\begin{titlepage}
\begin{center}

{\Large Chaotic strings and standard model parameters}

\vspace{1cm}

{\bf Christian Beck}\footnote{Permanent address:
School of Mathematical Sciences, Queen Mary,
University of London, Mile End Road, London E1 4NS, UK}

\vspace{0.5cm}

Isaac Newton Institute for Mathematical Sciences

University of Cambridge

20 Clarkson Road

Cambridge, CB3 0EH, UK

\end{center}

\vspace{1cm}

\abstract{We introduce so-called chaotic strings (coupled
1-dimensional noise strings underlying the Parisi-Wu approach of
stochastic quantization on a small scale) as a possible
amendment of ordinary string theories. These strings are
strongly self-interacting and exhibit strongest possible chao\-tic
behavior. Constraints on the vacuum energy of the strings fix a
certain discrete set of allowed string couplings. We provide
extensive numerical evidence that these string couplings
numerically coincide with running standard model coupling
constants, evaluated at energy scales given by the masses of the
known quarks, leptons and gauge bosons. Chaotic strings can thus
be used to provide a theoretical argument why certain standard
model parameters are realized in nature, others are not, assuming
that the {\em a priori} free standard model couplings evolve to the minima of
the effective potentials. The chaotic string spectrum correctly
reproduces the numerical values of the electroweak and strong
coupling constants with a precision of 4-5 digits, as well as the
(free) masses of the known quarks and leptons with a precision of
3-4 digits. Neutrino mass predictions consistent with
present experiments are obtained.
The $W$ boson mass also comes out correctly, and a
Higgs mass prediction is obtained.}

\vspace{1.3cm}

\end{titlepage}

\section{Introduction}

A fundamental problem of particle physics is the fact that there
are about 25 free fundamental constants which are not understood
on a theoretical basis. These constants are essentially the values
of the three coupling constants, the quark and lepton masses, the
$W$ and Higgs boson mass, and various mass mixing angles. An
explanation of the observed numerical values is ultimately
expected to come from a larger theory that embeds the standard
model. Prime candidates for this are superstring and $M$ theory
\cite{sustr1}--\cite{sustr3}. However, so far the predictive power
of these and other theories is not large enough to allow for
precise numerical predictions.

In this paper we will report on a numerical observation that may
shed more light on this problem. We have found that there is a
simple class of 1+1-dimensional strongly self-interacting discrete
field theories (called 'chaotic strings' in the following) that
have a remarkable property. The expectation of the vacuum energy
of these strings is minimized for string couplings that
numerically coincide with running standard model couplings $\alpha
(E)$, the energy $E$ being given by the masses of the known
quarks, leptons, and gauge bosons. Chaotic strings can thus be
used to provide theoretical arguments why certain standard model
parameters are realized in nature, others are not. We may assume
that the {\em a priori} free parameters evolve to the local minima
of the effective potentials generated by the chaotic strings. Out
of the many possible vacua, chaotic strings may select the
physically relevant vacuum of superstring theories.

The dynamics of the chaotic strings is discrete in both space and
time and exhibits strongest possible chaotic behaviour. It can be
regarded as a dynamics of vacuum fluctuations that can be used to
2nd-quantize other fields, for example ordinary standard model
fields, or ordinary strings, by dynamically generating the noise
of the Parisi-Wu approch of stochastic quantization \cite{stoch1, stoch2}
on a very
small scale. Mathematically, chaotic strings are
coupled map lattices \cite{kaneko} of diffusively coupled
Tchebyscheff maps $T_N$ of order N. It turns out that
there are six different relevant chaotic string theories
---similar to the six components
that make up M-theory in the moduli space of superstring theory
\cite{sustr3}. We will label these six chaotic string theories as
$3A,3B,2A,2B,2A^-,2B^-$. Here the first number denotes the index
$N$ of the Tchebyscheff polynomial and the letter A,B
distinguishes between the forward and backward coupling form. The
index $\,^-$ denotes anti-diffusive coupling. Though in principle
one can study these string theories for arbitrary $N$, for
stochastic quantization only the cases $N=2$ and $N=3$ yield
non-trivial behaviour in a first and second order perturbative
approach \cite{beck1, hilgers}.


Chaotic strings can be used to generate effective potentials for
self-inter\-acting scalar (dilaton-like) fields. They may be
effectively used to break supersymmetry via second quantization effects
(though a concrete embedding into superstring or $M$ theory is still a long
theoretical way to go, if any such embedding exists at all).
Assuming that the {\em a priori} free standard model couplings
evolve to the minima of the effective potentials generated by the
chaotic strings, one can obtain a large number of very precise
predictions. The smallest stable zeros of the expectation of the
interaction energy of the chaotic $3A$ and $3B$ strings are
numerically observed to coincide with the running electroweak
couplings at the smallest fermionic mass scales. Inverting the
argument, the chaotic 3A string can be used to theoretically
predict that the low-energy limit of the fine structure constant
has the numerical value
$\alpha_{el}(0)=0.0072979(17)=1/137.03(3)$, to be compared with
the experimental value $1/137.036$. The $3B$ string predicts that
the effective electroweak mixing angle is numerically given by
$\bar{s}_l^2=sin^2\theta_{eff}^{lept}=0.23177(7)$, in perfect
agreement with the experimental measurements at LEP, which yield
the value $\bar{s}_l^2=0.23185(23)$ \cite{pada}. The smallest
stable zeros of the interaction energy of the $N=2$ strings are
observed to coincide with strong couplings at the smallest bosonic
mass scales. In particular, the smallest stable zero of the
interaction energy of the $2A$ string yields a very precise
prediction of the strong coupling at the $W$ mass scale, which, if
evolved to the $Z^0$ scale, corresponds to the prediction
$\alpha_s(m_{Z^0})=0.117804(12)$. The current experimentally
measured value is $\alpha_s(m_{Z^0})=0.1185(20)$ \cite{pada}.

Besides the coupling strengths of the three interactions, also the
fermion mass spectrum can be obtained with high precision from
chaotic strings. Here the expectation of the self energy of the
chaotic strings is the relevant observable. One observes a large
number of string couplings that locally minimize the self energy
and at the same time numerically coincide with various running
electroweak, strong, Yukawa and gravitational couplings, evaluated
at the mass scales of the higher fermion families. The highest
precision predictions for fermion masses comes from the self
energy of the $2A$ and $2B$ strings, which is observed to exhibit
minima for string couplings that coincide with gravitational and
Yukawa couplings of all known fermions. The spectrum of these
strings yields the free masses of the six quarks as $m_u=5.07(1)$
MeV, $m_d=9.35(1)$ MeV, $m_s=164.4(2)$ MeV, $m_c=1.259(4)$ GeV,
$m_b=4.22(2)$ GeV and $m_t=164.5(2)$ GeV. Note that a free top
mass prediction of 164.5(2) GeV corresponds to a top pole mass
prediction of 174.4(3) GeV, in very good agreement with the
experimentally measured value $M_t=174.3 \pm 5.1$ GeV. The masses
of the charged leptons come out as $m_e=0.5117(8)$ MeV,
$m_\mu=105.6(3)$ MeV and $m_\tau=1.782(7)$ GeV. All these
theoretically obtained values of fermion masses are in perfect
agreement with experimental measurements. To the best of our
knowledge, there is no other theoretical model that has achieved
theoretical predictions of similar precision. Chaotic strings also
provide evidence for massive neutrinos, and yield
concrete predictions for the masses of the neutrino mass eigenstates
$\nu_1,\nu_2,\nu_3$. These are $m_{\nu_1}=1.452(3)\cdot 10^{-5}$ eV,
$m_{\nu_2}=2.574(3)\cdot 10^{-3}$ eV, $m_{\nu_3}=4.92(1)\cdot 10^{-2}$ eV.

Not only fermion masses, but also boson masses can be obtained
from chaotic strings. The 2A string correctly reproduces the
masses of the $W$ and $Z$ boson, and a suitable interpretation of
the $2B^-$ string dynamics provides evidence for the existence of
a scalar particle of mass $m_H=154.4(5)$ GeV, which could be
identified with the Higgs particle. The latter mass prediction is
slightly larger than supersymmetric expectations but well within
the experimental bounds based on the ordinary standard model. We
also obtain estimates of the lightest glueball masses, which are
consistent with estimates from lattice QCD.

This paper is organized as follows. In section 2 we introduce
chaotic strings. In section 3 we write down (formal) potentials
for the strings and define two types of vacuum energies, the
interaction energy $W(a)$ and the self-energy $V(a)$ of the
chaotic string. Essentially $W(a)$ fixes coupling constants and
$V(a)$ masses and mass mixing angles. In section 4 we introduce a
physical interpretation of the chaotic string dynamics in terms of
rapidly fluctuating virtual momenta. In sections 5 we will
consider the analogue of an Einstein equation which makes
arbitrary couplings evolve to the stable zeros of the interaction
energy of the strings. We will provide numerical evidence that the
smallest stable zeros of the $N=3$ strings numerically coincide
with the electroweak coupling constants at the smallest fermionic
mass scales. We then turn to $N=2$ strings and provide numerical
evidence that the smallest stable zeros coincide with strong
couplings at the smallest bosonic mass scales. In section
6 we will consider a suitable scalar field equation and provide
evidence that local minima of the self-energy of the chaotic
strings correctly reproduce the masses of all higher fermion
families. We extract the fermion masses from the self energy of
the $2A$, $2B$ strings, as well as the boson masses from the
$2A^-$, $2B^-$ strings. Our concluding remarks are given in
section 7.

\section{Chaotic Strings}

Chaotic strings are motivated by the assumption that there is a
small-scale dynamics underlying the noise fields of the Parisi-Wu
approach of stochastic quantization \cite{beck1}. Since in string theory
ordinary particles are believed to have string-like structure, it
is natural to assume that also the noise used for second
quantization in the Parisi-Wu approach may have such a string-like
structure on a very small scale.

Among the many models that can be chosen to generate a
deterministic chaotic noise string dynamics on a small scale
certain criteria should be applied to select a particular system.
First of all, for vanishing spatial coupling of the chaotic
`noise' one wants to have strongest possible random behavior with
least possible higher-order correlations, in order to be closest
to the Gaussian limit case (which corresponds to ordinary path
integrals on a large scale). This selects as a local dynamics
Tchebyscheff maps $T_N(x)$ of $N$-th order ($N\geq 2$)
\cite{beck3}. It is well known that Tchebyscheff maps satisfy a
Central Limit Theorem which guarantees the convergence to the
Wiener process (and hence to ordinary path integrals) if sums of
iterates are looked at from large scales \cite{bill, roep}. As
shown in \cite{beck3} Tschebyscheff maps have least higher-order
correlations among all systems conjugated to a Bernoulli shift,
and are in that sense closest to Gaussian white noise, though
being completely deterministic. A graph theoretical method for
this type of noise has been developed in \cite{beck3, new}.

Moreover, when spatially coupling the discrete chaotic dynamics
this should formally result from a Laplacian coupling rather than
some other coupling, since this is the most relevant coupling form
in quantum field and string theories. This leads to coupled map
lattices of the diffusive coupling form. The resulting coupled map
lattices can then be studied on lattices of arbitrary dimension
(as done in \cite{beck2}), but motivated by the fact that ordinary
strings are 1-dimensional objects we will focus in this paper on
the 1-dimensional case. One obtains a `chaotic string' defined by
\begin{equation}
\Phi_{n+1}^i=(1-a)T_N(\Phi_n^i) + s \frac{a}{2} (T_N^b
(\Phi_n^{i-1}) +T_N^b(\Phi_n^{i+1})). \label{dyn}
\end{equation}
$\Phi_n^i$ is a discrete chaotic noise field variable taking
continuous values on the interval $[-1,1]$. The initial values
$\Phi_0^i$ are randomly distributed in this interval. $i$ is a
spatial lattice coordinate and $n$ a discrete time coordinate (in
our case identified with the fictitious time of the Parisi-Wu
approach). $T_N$ denotes the $N$-th order Tchebyscheff polynomial.
In the following we will mainly study $T_2(\Phi)=2\Phi^2-1$ and
$T_3(\Phi)=4\Phi^3-3\Phi$. We consider both the positive and
negative Tchebyscheff polynomial $T_N^\pm (\Phi)=\pm T_N(\Phi)$,
but have suppressed the index $\pm$ in the above equation. The
variable $a$ is a coupling constant taking values in the interval
$[0,1]$. Since $a$ determines the strength of the Laplacian
coupling, $a^{-1}$ can be regarded as a kind of metric in the
1-dimensional string space indexed by i. $s$ is a sign variable
taking on the values $\pm1$. The choice $s =+ 1$ is called
`diffusive coupling', but for symmetry reasons it also makes sense
sense to study the choice $s=-1$, which we call `anti-diffusive
coupling'. The integer $b$ distinguishes between the forward and
backward coupling form \cite{beck2}, $b=1$ corresponds to forward
coupling ($T_N^1(\Phi):=T_N(\Phi))$, $b=0$ to backward coupling
($T_N^0(\Phi):=\Phi$). We consider periodic boundary conditions
and large lattices of size $i_{max}$.

One can easily check that for odd $N$ the choice of $s$ is
irrelevant (since odd Tchebyscheff maps satisfy $T_N(-\Phi
)=-T_N(\Phi )$), whereas for even $N$ the sign of $s$ is relevant
and a different dynamics arises. Hence, restricting ourselves to
$N=2$ and $N=3$, in total 6 different chaotic string theories
arise, characterized by
$(N,b,s)=(2,1,+1),(2,0,+1),(2,1,-1),(2,0,-1)$ and
$(N,b)=(3,1),(3,0)$. For easier notation, in the following we will
label these string theories as $2A,2B,2A^-,2B^-,3A,3B$,
respectively.

If the coupling $a$ is sufficiently
small, the chaotic variables $\Phi_n^i$ can be used to generate the noise
of the Parisi-Wu approach of stochastic quantization on a very
small scale. It can actually be shown that if chaotic Tchebyscheff
noise is coupled to a slowly varying dynamics, in first and second
order perturbation theory the cases $N>3$ do not yield
anything new compared to Gaussian white noise (see \cite{hilgers}
for more details). Hence the above six chaotic string theories
obtained for $N=2$ and $N=3$ are the most relevant ones to
consider, yielding non-trivial behaviour in leading order of
stochastic quantization.

As shown in \cite{beck1, beck2}, the chaotic noise string dynamics
(\ref{dyn}) formally originates from a 1-dimensional continuum
$\phi^{N+1}$-theory in the limit of infinite self-interaction
strength (see Appenidix A for deriving, as an example, the $3B$
string dynamics). In this sense, chaotic strings can also be
regarded as degenerated Higgs-like fields with infinite self-interaction
parameters, which are
constraint to a 1-dimensional space.

\section{Vacuum energy of chaotic strings}

Though the chaotic string dynamics is dissipative, one can
formally introduce potentials that generate the discrete time
evolution. For $a=0$ we may write
\begin{equation}
\Phi_{n+1}-\Phi_n=\pm T_N(\Phi_n)=-\frac{\partial}{\partial \Phi_n}V_\pm (\Phi_n).
\end{equation}
For $N=2$ the (formal) potential is given by
\begin{equation}
V^{(2)}_\pm (\Phi )= \pm
\left( -\frac{2}{3} \Phi^3 +\Phi \right)
+\frac{1}{2} \Phi^2+C, \label{v2}
\end{equation}
for $N=3$ by
\begin{equation}
V^{(3)}_\pm (\Phi )=\pm \left( -\Phi^4+\frac{3}{2}
\Phi^2 \right) +\frac{1}{2} \Phi^2 +C . \label{v3}
\end{equation}
Here $C$ is an arbitrary constant. The uncoupled case $a=0$ is
completely understood. The dynamics is ergodic and mixing. Any
expectation of an observable $A(\Phi)$ can be calculated as
\begin{equation}
\langle A \rangle = \int_{-1}^1 d\phi \; \rho (\phi)  A(\phi ),
\end{equation}
where
\begin{equation}
\rho (\phi )=\frac{1}{\pi \sqrt{1-\phi^2}} , \label{berlin}
\end{equation}
is the natural invariant density describing the probability
distribution of iterates of Tchebyscheff maps (see, e.g.\
\cite{BS}). In the formalism of nonextensive
statistical mechanics \cite{tsallis}, this probability
density can be regarded as a
generalized canonical distribution with entropic index $q=3$,
or as an escort distribution with index $q=-1$.

If a spatial coupling $a$ is introduced, things become much more
complicated, and the invariant 1-point density deviates from the
simple form (\ref{berlin}). A spatial coupling is formally
generated by the interaction potential $aW_\pm (\Phi ,
\Psi )$, with
\begin{equation}
W_\pm (\Phi , \Psi ) = \frac{1}{4} (\Phi \pm \Psi )^2 +C. \label{w}
\end{equation}
Here $\Phi$ and $\Psi$ are neighboured noise field variables on the lattice.
One has
\begin{equation}
-\frac{\partial}{\partial \Phi^i}W_\pm (\Phi^i , \Phi^{i+1})
-\frac{\partial}{\partial \Phi^i}W_\pm (\Phi^i , \Phi^{i-1})
= \pm \frac{1}{2} \Phi^{i+1}-\Phi^i \pm \frac{1}{2} \Phi^{i-1}.
\end{equation}
This generates diffusive $(+)$, respectively anti-diffusive $(-)$ coupling.
Anti-diffusive coupling can equivalently be obtained by keeping
$W_-$ but replacing
$T_N\to -T_N$ at odd lattice sites.
The coupled map dynamics (\ref{dyn})
is obtained by letting the action of
$V$ and $W$ alternate in discrete time $n$, then regarding
the two time steps as one. For the backward coupling form $b=0$,
the action of the potentials also alternates in discrete space $i$.

The expectations of the potentials $V$ and $W$ yield two types of
vacuum energies $V(a):=\langle V_\pm^{(N)}(\Phi^i ) \rangle$ (the
self energy) and $W(a):=\langle W_\pm (\Phi^i ,\Phi^{i+1} )
\rangle$ (the interaction energy). Here $\langle \cdots \rangle$
denotes the expectation with respect to the coupled chaotic
dynamics. Numerically, any such expectation can be determined by
averaging over all $i$ and $n$ for random initial conditions
$\Phi^i_0\in [-1,1]$, omitting the first few transients. Note that
in the stochastic quantization approach the chaotic
noise is used for 2nd quantization of standard model fields (or
ordinary strings) via the Parisi-Wu approach \cite{beck1}. Hence generally
expectations with respect to the chaotic dynamics correspond to
expectations with respect to 2nd quantization. The expectation of
the vacuum energy of the string, given by the above functions
$W(a)$ and $V(a)$, depends on the coupling $a$ in a non-trivial
way. Moreover, it also depends on the integers $N,b,s$ that define
the chaotic string theory.

Since negative and positive Tchebyscheff maps essentially generate
the same dynamics, up to a sign, any physically relevant
observable should be invariant under the transformation $T_N\to
-T_N$. The vacuum energies $V(a)$ and $W(a)$ of the various
strings exhibit full symmetry
under the transformation $T_N \to -T_N$ (respectively $s \to -s$) if the
additive constant $C$ is chosen to be
\begin{equation}
C= -\frac{1}{2} \langle \Phi^2 \rangle.
\end{equation}
For that choice of $C$, the expectations of $V_+$ and $V_-$ as well
as those of $W_-$ and $W_+$ are the same, up
to a sign.

Choosing (by convention) the
$+$ sign one obtains from eq.~(\ref{v2}), (\ref{v3}) and (\ref{w})
for the expectations of the potentials
\begin{eqnarray}
V^{(2)}(a) &=& -\frac{2}{3} \langle \Phi^3 \rangle +\langle \Phi
\rangle \\ V^{(3)}(a) &=&  -\langle \Phi^4 \rangle +\frac{3}{2}
\langle \Phi^2 \rangle ,
\end{eqnarray}
and
\begin{equation}
W(a) =\frac{1}{2} \langle \Phi^{i} \Phi^{i+1} \rangle .
\end{equation}

The above $\pm$ symmetry can actually be used to cancel unwanted vacuum
energy and to avoid problems with the cosmological constant. If
one assumes that strings with both $T_N$ and $-T_N$ are physically
relevant, the two contributions $V(a)$ and $-V(a)$ (respectively
$W(a)$ and $-W(a)$) may simply add up to zero. This reminds us of
similarly good effects that supersymmetric partners have in
ordinary quantum field and string theories. As a working
hypothesis, we will thus formally associate the above symmetry
with supersymmetry in the chaotic noise space.

Similarly as for ordinary strings it also makes sense to consider
certain conditions of constraints for the chaotic string. For
ordinary strings (for example bosonic strings in covariant gauge
\cite{sustr1}) one has the condition of constraint that the energy
momentum tensor should vanish. The first diagonal component of the
energy momentum tensor is an energy density. For chaotic strings,
the evolution in space $i$ is governed by the potential $W_\pm
(\Phi, \Psi)$ and the corresponding expectation of the energy
density is $\pm W(a)$. We should thus impose the condition of
constraint that $W(a)$ should vanish for physically observable
states. Moreover, the evolution in fictitious time $n$ is governed
by the self-interacting potential $V^{(N)} (\Phi )$. This
potential generates a shift of information, since the Tchebyscheff
maps $T_N$ are conjugated to a Bernoulli shift of $N$ symbols.
Hence $V(a)$ can be regarded as the expectation of a kind of
information potential or entropy function, which, motivated by
thermodynamics, should be extremized for physically observable
states. Note that the action of $V$ and $W$ alternates in $n$ and
$i$ direction. Both types of vacuum energies describe different
relevant observables of the chaotic string and are of equal
importance.

\section{Physical interpretation of the chaotic string dynamics in
terms of vacuum fluctuations}

In order to
construct a link to standard model phenomenology it is useful to
introduce a simple physical interpretation of the chaotic string
dynamics in terms of fluctuating virtual momenta.
Suppose we regard $\Phi_n^i$ to be a fluctuating virtual momentum
component associated with a hypothetical particle $i$ at
(fictitious) time $n$ that lives in the constraint 1-dimensional
string space. Then neighboured particles $i$ and $i-1$ exchange
momenta due to the diffusive coupling. A more detailed physical
interpretation would be that at each time step $n$ a
particle-antiparticle pair $f_1,\bar{f}_2$ is being created in cell $i$ by the
field energy of the self-interacting potential. In units of some
arbitrary energy scale $p_{max}$, the particle has momentum
$\Phi_n^i$, the antiparticle momentum $-\Phi_n^i$. They interact
with particles in neighboured cells by exchange of a
(hypothetical) gauge boson $B_2$, then they annihilate into another
boson $B_1$ and the next
chaotic vacuum fluctuation (the next creation of a
particle-antiparticle pair) takes place. This can be symbollically
described by the Feynman graph in Fig.~1. Actually, the graph
continues ad infinitum in time and space and could thus be called
a `Feynman web', since it describes an extended spatio-temporal
interaction state of the string, to which we have given a standard
model-like interpretation. The important point is that in this
interpretation $a$ is a (hypothetical) standard model coupling
constant, since it describes the strength of momentum exchange of
neighboured particles. At the same time, $a$ can also be regarded
as an inverse metric in the 1-dimensional string space, since it
determines the strength of the Laplacian coupling.

It is well known that standard model interaction strengths
actually depend on the relevant energy scale $E$. We have the
running electroweak and strong coupling constants.
What should we now take for the energy (or temperature)
$E$ of the chaotic string? {\em A priori} this is unknown. However, we
will present extensive numerical evidence that the constraints on
the vacuum energy of the chaotic string fix certain discrete
string couplings $a_i$, and these string couplings are numerically
observed to coincide with running standard model couplings, the
energy (or temperature)
being given by
\begin{equation}
E= \frac{1}{2} N (m_{B_1}+m_{f_1}+m_{f_2}). \label{katharina}
\end{equation}
Here $N$ is the index of the chaotic string theory considered, and
$m_{B_1}, m_{f_1},m_{f_2}$ denote the masses of the particles
involved in the Feynman web interpretation. The surprising
observation is that rather than yielding just some unknown exotic
physics, the chaotic string spectrum appears to reproduce the
masses and coupling constants of the known quarks, leptons and
gauge bosons of the standard model (plus possibly more)
\footnote{If the presence of black holes is assumed, one can also
relate $E$ to the Hawking temperature $kT_H\sim  \frac{1}{GM}$,
thus connecting very large black hole masses $M$ with very small
temperatures $kT_H$.}.

Formula (\ref{katharina}) formally  reminds us of the energy
levels $E_N=\frac{N}{2} \hbar \omega$ of a quantum mechanical
harmonic oscillator, with low-energy levels ($N=2,3$) given by the
masses of the standard model particles. In the Feynman web
interpretation of Fig.~1, the formula is plausible. We expect the
process of Fig.~1 to be possible as soon as the energy per cell
$i$ is of the order $m_{B_1}+m_{f_1}+m_{f_2}$. The boson $B_2$ is
virtual and does not contribute to the energy scale. The factor
$N$ can be understood as a multiplicity factor counting the degrees
of freedom. Given some value
$\Phi_n^i$ of the momentum in cell $i$, there are $N$ different
pre-images $T_N^{-1}(\Phi_n^i )$ how this value of the momentum
can be achieved. All these different channels contribute to the
energy scale.


\section{Zeros of the interaction energy $W(a)$}

\subsection{Analogue of the Einstein field equations}

Let us now suppose that {\em a priori} arbitrary standard model
couplings $\alpha$ are possible. Assume that these couplings are
at the same time coupling constants $a$ in the chaotic string
space. In fact, we may then regard $\alpha^{-1}=a^{-1}$ as a kind
of metric in the 1-dimensional string space. A stochastically
quantized Einstein equation for this 1-dimensional (trivial)
metric in the 1-dimensional string space would then have the form
\begin{equation}
\left( \frac{1}{a} \right)^{\dot{\,}} = -\frac{1}{a^2} \dot{a}=
T_{00}+noise.
\end{equation}
Here $T_{00}$ is the energy density of the vacuum as created by
the interaction energy $W$ of the chaotic strings, and $t$
is the fictitious time of the Parisi-Wu approach, used
for second quantization. Choosing
$T_{00}=-\Gamma \cdot W$, where $\Gamma$ is a positive constant,
one ends up with
\begin{equation}
\dot{a} =a^2 \Gamma W(a ) +noise. \label{evo}
\end{equation}
The fictitious time $t$ of the Parisi-Wu approach has dimension
$energy^{-2}$, hence if $W(a)$ is defined to be dimensionless
then the constant $\Gamma$ should have dimension
$energy^2$, and we could for example choose $\Gamma\sim G^{-1}$ or
$\Gamma \sim a^{-2}G^{-1}$, where $G$ is the gravitational constant.

Apparently all zeros of the function $W(a)$ describe a stationary
state for possible standard model couplings, but only those zeros
where $W(a)$ has locally negative slope decribe a {\em stable}
stationary state. Arbitrary initial couplings $a$ will evolve to
these stable states (Fig.~2). Of course the stability properties
depend on the choice of the sign of the constant $\Gamma$. This
choice breaks the initial symmetry of the problem. Let us assume
that for our world $\Gamma >0$ is relevant and label the stable
zeros of $W(a)$ (i.e. those with negative slope) by $a_i^{(Nb)}$,
$i=1,2,\ldots$. Here $N$ is the index of the string theory
considered, and $b=1,0$ (or A,B) distinguishes between forward and
backward coupling. We will now describe our numerical findings in
detail.

\subsection{
The 3A string---electric interaction strengths of electrons and
$d$-quarks} Fig.~3 shows the interaction energy  $W(a)=\frac{1}{2} \langle
\Phi_n^i \Phi_n^{i+1} \rangle$ of the chaotic $3A$ string. We
observe the following stable zeros in the low-coupling region
$a\in [0,0.018]$:
\begin{eqnarray*}
a_1^{(3A)}&=&0.0008164(8)     \\ a_2^{(3A)}&=&0.0073038(17)
\end{eqnarray*}
The statistical error is estimated by repeating the iteration of the
coupled map lattice (\ref{dyn}) for different
random initial conditions $\Phi_0^i \in [-1,1]$.
We used coupled map lattices of size 10000
with periodic boundary conditions,
typical iteration times over which the product
$\Phi_n^i \Phi_n^{i+1}$ was averaged were $n_{max}=10^7$.

Remarkably, the zero $a_2^{(3A)}$ appears to approximately
coincide with the fine structure constant $\alpha_{el} \approx
1/137$. To construct a suitable Feynman web interpretation, let us
choose in Fig.~1 for $B_1$ any massless boson and $B_2=\gamma$,
$f_1=e^-,\bar{f}_2=e^+$. The relevant energy scale underlying this
Feynman web is given by $E=(3/2)(m_\gamma+2m_e)=3m_e$, according
to eq.~(\ref{katharina}). Hence our standard model interpretation
of the string state described by $a_2^{(3A)}$ suggests the
numerical identity
\begin{equation}
a_2^{(3A)}=\alpha_{el}(3m_e).
\end{equation}
For a precise numerical comparison
let us estimate the running electromagnetic coupling at this energy scale.
We may use the 1st-order QED formula
\begin{equation}
\alpha_{el} (E) = \alpha_{el} (0) \left\{ 1+ \frac{2\alpha_{el} (0)}{\pi} \sum_i f_i
\int_0^1dx\; x(1-x)  \log \left( 1+ \frac{E^2}{m_i^2}x(1-x)\right)
\right\}, \label{runael}
\end{equation}
The sum is over all charged elementary particles, $m_i$ denotes
their (free) masses, and $f_i$ are charge factors given by 1 for
$e,\mu ,\tau$-leptons, $\frac{4}{3}$ for $u,c,t$-quarks and
$\frac{1}{3}$ for $d,s,b$-quarks. Using this formula, we get
$\alpha_{el}(3m_e) =0.007303$, to be compared with
$a_2^{(3A)}=0.0073038(17)$. There is excellent agreement.

Next, we notice that the zero $a_1^{(3A)}$ has approximately the value
$\frac{1}{9} \alpha_{el}$. This could mean that the chaotic 3A
string also has a mode that provides evidence for electrically
interacting $d$-quarks. Our interpretation is
\begin{equation}
a_1^{(3A)} =\alpha_{el}^d (3 m_d)=\frac{1}{9} \alpha_{el} (3 m_d),
\end{equation}
where $\alpha_{el}^d=\frac{1}{9}\alpha_{el}$ denotes the
electromagnetic interaction strength of $d$-quarks. In the Feynman
web interpretation, we have $B_1$ $massless$, $B_2=\gamma$,
$f_1=d,f_2=\bar{d}$. Formula (\ref{runael}), as an estimate,
yields for $m_d= 9$ MeV the value $\alpha_{el}(3m_d)= 0.007349$,
which coincides very well with $9a_1^{(3A)}=0.007348(7)$. The
value $9a_1^{(3A)}$ actually translates to the energy scale
$E_d=(26.0 \pm 6.4)$ MeV. This yields $m_d = \frac{1}{3}E_d=(8.7
\pm 2.1)$ MeV, which coincides with estimates of the
$\overline{MS}$ current quark mass of the $d$ quark at the
proton mass renormalization scale \cite{pada}.


\subsection{The 3B string ---weak interaction strengths
of neutrinos and $u$-quarks}

The 3B string is a
system with backward coupling. For backward coupling the action of
the potentials $V$ and $W$ not only alternates
in time $n$ but also in discrete space $i$.
Hence one may conjecture that this coupling form generally
describes mixed states of two particles.

The interaction energy $W(a)$ of the 3B string is plotted in
Fig.~4. In the low-coupling region $a\in[0,0.018]$ we observe the
following stable zeros of $W(a)$:
\begin{eqnarray*}
a_1^{(3B)} &=& 0.0018012(4) \\ a_2^{(3B)} &=& 0.017550(1)
\end{eqnarray*}
If our approach is consistent, we might be able to find an
interpretation of $a_1^{(3B)}$ and $a_2^{(3B)}$ indicating the
existence of $u$-quarks and neutrinos.

Let us start with $a_2^{(3B)}$. For left-handed neutrinos, the weak
coupling due to the exchange of $Z^0$-bosons is given by
\begin{equation}
\alpha_{weak}^{\nu_L}=\alpha_{el} \frac{1}{4 \sin^2 \theta_W \cos^2 \theta_W}.
\end{equation}
Here $\theta_W$ is the weak mixing angle. In the following we will
treat $\sin^2 \theta_W $ as an effective constant, and regard
$\alpha_{el}$ as the running electromagnetic coupling. Other
renormalization schemes are also possible, but yield only minor
numerical differences. Experimentally, at LEP the effective weak mixing
angle is measured as $\sin^2 \theta_W=0.2318(2)$
\cite{pada}. Assuming that in addition to the left-handed neutrino
interacting weakly there is an electron interacting electrically,
the two interaction processes can add up independently if the
electron is right-handed, since right-handed electrons cannot
interact with left-handed neutrinos. Hence a possible standard
model interpretation of the zero $a_2^{(3B)}$ would be
\begin{equation}
a_2^{(3B)}=\alpha_{el}(3 m_e) +\alpha_{weak}^{\nu_L} (3 m_{\nu_e} )
=a_2^{(3A)}+ \alpha_{el} (3m_{\nu_e}) \frac{1}{4\sin^2 \theta_W \cos^2
\theta_W} \label{61}
\end{equation}
In the Feynman web interpretation of Fig.~1, $B_2=Z^0$,
$f_1=\nu_L$, $\bar{f}_2=\bar{\nu_L}$ in addition to the process already
described by $a_2^{(3A)}$. $B_1$ can again be any massless boson. Putting in
the experimentally measured value of $\sin^2\theta_W=0.2318$, we
obtain for the right-hand side of eq.~(\ref{61}) the value
0.01755, which coincides perfectly with the observed string
coupling $a_2^{(3B)}=0.017550$.

Next, let us interpret $a_1^{(3B)}$. In analogy to the joint
appearance of $\nu$ and $e$, we should also expect to find a
weakly interacting $u$-quark, together with a $d$-quark
interacting electrically. Clearly, the $u$-quark could also
interact electrically, but for symmetry reasons we expect the pair
$(u,d)$ to interact in a similar way as $(\nu ,e)$. A right-handed
$u$-quark interacts weakly with the coupling
\begin{equation}
\alpha_{weak}^{u_R} = \frac{4}{9} \alpha_{el} \frac{\sin^2 \theta_W}{
\cos^2 \theta_W}.
\end{equation}
Adding up the electrical interaction strength of a $d$-quark, a
natural interpretation, quite similar to that of the zero
$a_2^{(3B)}$, is
\begin{equation}
a_1^{(3B)}=\alpha_{el}^{d} (3 m_d) + \alpha_{weak}^{u_R}(3 m_u)
 = a_1^{(3A)}+ \frac{4}{9} \alpha_{el} (3 m_u)
\frac{\sin^2 \theta_W}{\cos^2 \theta_W}  \label{america}
\end{equation}
The Feynman web interpretation of this string state is $B_1\;
massless$, $B_2=Z^0$, $f_1=u_R$, $f_2=\bar{u_R}$ in addition to
the process underlying $a_1^{(3A)}$. Numerically, taking $\sin^2
\theta_W=0.2318$ and evaluating the running $\alpha_{el}$ using
$m_u=5$ MeV, we obtain for the right-hand side of
eq.~(\ref{america}) 0.001800, which should be compared with
$a_2^{(3B)}=0.001801$. Again we have perfect agreement within the
first 4 digits. It is remarkable that the same universal effective
value $\sin^2 \theta_W =0.2318$ can be used consistently for both
leptons (couplings $a_2^{(3A)}$, $a_2^{(3B)})$ and quarks
(couplings $a_1^{(3A)}$, $a_1^{(3B)}$). If, on the other hand, we
assume that $\sin^2 \theta_W$ is fixed by $a_2^{(3B)}$ to be
0.2318, the running electric coupling can be used to extract the
energy scale $E$ from $a_1^{(3B)}$. We obtain $E=3 m_u=(21.3
\pm 9)$ MeV, meaning $m_u = (7.1 \pm 3.0)$ MeV.

Note that generally the backward coupling form of the $N=3$
strings seems to describe
a spinless state (formed by $e_R$ and $\nu_L$, respectively $d_L$
and $u_R$), whereas the forward coupling form just describes one
particle species with non-zero spin ($e$ or $d$). A similar statement will
turn out to hold for
the $N=2$ theories, replacing fermions by bosons.


\subsection{High-precision prediction of the electroweak parameters}

It appears that the smallest stable zeros of the $N=3$ strings coincide with the electroweak
coupling strengths at the mass scales of the lightest fermions
$d,u,e,\nu_e$. We can thus actually invert the formulas and give a rather
precise prediction of the electroweak parameters.
The general idea underlying this is that at some early stage of the universe
(possibly in a pre-big bang scenario)
the electroweak
couplings were not fixed but were following the
evolution equation (\ref{evo}), being finally fixed
as stable inverse metrics in the string space. 
The chaotic string dynamics may still
persist today and stabilize the observed parameter values. The observed
low-energy limit of the fine structure
constant follows from $a_2^{(3A)}$ and formula (\ref{runael}) as
\begin{eqnarray}
\alpha_{el}(0)&=&\! a_2^{(3A)} \! \left( 1-\frac{2a_2^{(3A)}}\pi \sum_i f_i
\int_0^1dx \; x(1-x) \log \left( 1 + \frac{9 m_e^2}{m_i^2}x(1-x)
\right)\right)  \nonumber \\ &=& 0.0072979(17)
=1/137.03(3).
\end{eqnarray}
Here we used the known value of the electron mass and neglected
terms of order $\alpha^3$, which are much smaller than the
statistical error. Within the error bounds, our result agrees very
well with the experimentally measured value $\alpha
(0)=0.00729735=1/137.036$ (the experimentally measured number is
known to quite precisely coincide with $\frac{1}{137}\cdot \cos
\frac{\pi}{137} =0.00729735$ \cite{gil}).

Next, the effective weak mixing angle can be predicted from
eq.~(\ref{61}). Since $m_{\nu_e}\approx 0$ one obtains
\begin{equation}
\sin^2 \theta_W =\frac{1}{2} \left( 1- \sqrt{1- \frac{\alpha
(0)}{a_2^{(3B)}-a_2^{(3A)}}} \right) \label{62}
\end{equation}
This yields
\begin{equation}
\sin^2 \theta_W =0.23177(7). \label{63}
\end{equation}
The value perfectly coincides with experimental measurements
\cite{pada}. In fact, the prediction is much more precise
than the supersymmetric GUT model prediction 0.234(2). Within the
error bars, we actually do get the same effective Weinberg angle
for quarks and leptons using either $a_1^{(3B)}$ or $a_2^{(3B)}$.

Any high-precision result on $\sin^2 \theta_W$ can also be used to
estimate gauge and Higgs boson masses from radiative corrections.
For example we may use the results of a calculation by Degrassi et
al. \cite{degra}.
Our Feynman web interpretation of the zero $a_2^{(2B)}$ suggests to
regard our $\sin^2 \theta_W$ as the effective mixing parameter
$\bar{s_l}^2=\sin^2\theta_{eff}^{lept}$ for $Z^0$-lepton coupling. Table 1 in
\cite{degra} translates the value $0.23177(7)$ to $m_W =80.35(1)$
GeV and yields a Higgs mass of the order of twice the $W$ mass.

Note that there is excellent agreement of the weak mixing angle
obtained from the chaotic 2B string with the experimental
measurements at LEP. At LEP the effective weak mixing angle
$\bar{s_l}^2$ is measured as 0.23185(23), from this the Higgs mass
is estimated as $166^{+270}_{-95}$ GeV and the W-mass as
$80.35(6)$ GeV \cite{pada}.

\subsection{The 2A string
---strong interaction strength at the $W$-mass scale}
If electroweak coupling strengths are fixed by suitable zeros of
the interaction energy of chaotic strings, then something similar
should also be the case for the strong coupling strength
$\alpha_s$. Let us now look at strings with $N=2$.
Fig.~5 shows the
interaction energy $W(a)$ of the 2A string. Only one stable zero
is observed:
\begin{equation}
a_1^{(2A)} = 0.120093(3) \label{strong}
\end{equation}
We notice that it numerically seems to coincide with the strong
coupling constant $\alpha_s$ at the $W$- or $Z$ mass scale,
which is experimentally
measured as $\alpha_s (m_Z)=0.1185(20)$ \cite{pada}.

For symmetry reasons, it seems plausible that if the $N=3$ strings
fix the electroweak couplings at the smallest fermionic mass
scales, then $N=2$ strings could fix the strong couplings at the
smallest bosonic mass scales. The lightest massive gauge boson is
indeed the $W^\pm$. Hence our Feynman web interpretation
associated with $a_1^{(2A)}$ is $B_1=W^\pm, B_2=g$ (gluon), $f_1=u$,
$\bar{f_2}=\bar{d}$ (respectively $f_1=d, \bar{f_2}=\bar{u}$), and since $N=2$ formula (\ref{katharina}) implies
\begin{equation}
a_1^{(2A)}=\alpha_s(m_W+m_u+m_d) \approx \alpha_s (m_W).
\label{Franziska}
\end{equation}
Since the $W$-mass is known with high precision, eq.~(\ref{Franziska})
yields
quite a precise prediction for the strong coupling $\alpha_s$. We can
evolve it to arbitrary energy scales,
using the well known perturbative results from QCD (see Appendix B).
In 3rd-order perturbation theory eq.~(\ref{Franziska}) is equivalent to
an effective QCD scale parameter
\begin{equation}
\Lambda^{(5)} =0.20608(14) GeV,
\end{equation}
neglecting higher-order terms. Here we use for the $W$-mass the
value measured at LEP, $m_W=80.35(6)$ GeV \cite{pada}. At $E=m_{Z}=91.188(2)$
GeV eq.~(\ref{runas}) in Appendix B yields
\begin{equation}
\alpha_s (m_{Z^0})=0.117804(12).
\end{equation}
This prediction of $\alpha_s$ from the zero of the chaotic 2A string
is clearly consistent with the
experimentally measured value 0.1185(20)
and in fact much more precise than current experiments can verify.

To evolve $\alpha_s(E)$ to energies $E>M_t$ or $E<M_b$ one has to
take into account quark threshold effects. In chapter 6 we will
obtain from the self energy of the chaotic strings the
free\footnote{see Appendix B for the definition of 'free' quark
masses} bottom and top masses as $m_b=4.23(1)$ GeV and $m_t=
164.5(2)$ GeV. The corresponding pole masses are $M_b=4.87(2)$ GeV
and $M_t=174.3(3)$ GeV (formula (\ref{pole}) in Appendix B).
Matching the running $\alpha_s (E)$ in a continuous way at $M_t$
and $M_b$ \cite{marciano}, we obtain from the above value
$\Lambda^{(5)}=0.20608(14)$ GeV the values
$\Lambda^{(6)}=0.08705(6)$ GeV and $\Lambda^{(4)} =0.28913(17)$
GeV. Slight numerical differences can arise from the way the quark
thresholds are treated.

\subsection{The 2B string ---the lightest scalar glueball}
The interaction energy of the 2B string is shown in Fig.~6. $W(a)$
has only one non-trivial zero
\begin{equation}
a_1^{(2B)}=0.3145(1).
\end{equation}
It has negative slope, so it should describe an observable stable state.
One possibility is to interpret
this as a strong coupling
at the lightest glueball mass scale. The lightest scalar glueball has spin $J^{PC}=0^{++}$
and is denoted by $gg^{0++}$ in the following. If we choose the Feynman web
interpretation
$B_1=gg^{0++}, B_2=g$,
$f_1=u$, $\bar{f_2}=\bar{u}$), meaning 
\begin{equation}
a_1^{(2B)}=\alpha_s (m_{gg^{0++}}+2m_u) \approx \alpha_s(m_{gg^{0++}}),
\label{glueball}
\end{equation}
then indeed the 2B string describes two bosons (two
gluons forming a glueball), similar to the 3B
string, which described two fermions. In both cases a spin 0 state is formed.
In lattice gauge
calculations including dynamical fermions the smallest scalar glueball
mass is estimated as $m_{gg^{0++}}=(1.74 \pm 0.07)$ GeV \cite{100} and
at this energy the running strong coupling constant is
experimentally measured to be $\alpha_s =0.32 \pm 0.05$
\cite{peach}. This clearly is consistent with the observed value
of $a_1^{(2B)}$.

We can estimate the lightest glueball mass predicted by the 2B
string using once again the 3rd-order QCD formula (\ref{runas}) in
Appendix B. The value $\alpha_s(m_{gg^{0++}}+2m_u)=0.3145(1)$ then
translates to $m_{gg^{0++}}+2m_u= 1.812(2)$ GeV.

\subsection{The 2A$\,^-$ and 2B$\,^-$ strings --- towards a Higgs mass prediction}
Two chaotic string theories are still remaining, namely those with $N=2$ and
antidiffusive coupling.
The interaction energies $W(a)$ of the 2A$\,^-$ and 2B$\,^-$ strings are shown in
Fig.~7 and 8. Let us now try to find
a suitable Feynman web
interpretation for the observed smallest stable zeros $a_1^{(2A^-)}=0.1758(1)$
and $a_1^{(2B^-)}=0.095370(1)$ of these strings.

Again let us be guided by symmetry considerations.
We saw that the smallest stable zero of the 2A string described a boson with non-zero spin (the
lightest massive gauge boson $W^\pm$) and the smallest stable zero of the 2B string a boson without spin
(the lightest scalar glueball). Thus it seems reasonable to assume that the smallest
stable zero of the 2A$\,^-$ string describes yet another bosonic particle with non-zero spin, possibly the
lightest glueball with spin $J^{PC}=2^{++}$, and the smallest zero of the 2B$\,^-$ string yet another
bosonic particle with spin 0, possibly the lightest Higgs boson.

Let us start with $a_1^{(2A^-)}$. Our Feynman web interpretation of this string state is
$B_1=gg^{2++}, B_2=g, f_1=q, \bar{f}_2=\bar{q}$, where
$q, \bar{q}$ are suitable quarks.
From the strong coupling interpretation
\begin{equation}
a_1^{(2A^-)}=\alpha_s (E_1)
\end{equation}
the energy $E_1=m_{gg^{2++}}+2m_q$ can again be determined from
the QCD formula (\ref{runas}), using our previously determined
$\Lambda^{(5)}$. One obtains $E_1=10.45(3)$ GeV. In lattice gauge
calculations the mass of the lightest $2^{++}$ glueball is
estimated as $2.23(31)$ GeV \cite{pada}. We thus get the correct
order of magnitude of the $2^{++}$ glueball mass if we assume that
the quarks in the Feynman web interpretation are bottom quarks. In
this case a glueball mass  $m_{gg2^{++}}=(10.45(3)-2\cdot
4.23(1))$ GeV =1.99(4) GeV is predicted, using $m_b=4.23(1)$ GeV.

Next, let us consider $a_1^{(2B^-)}$. This zero is even more interesting, since
it may provide evidence for the Higgs particle. Our Feynman web
interpretation is $B_1=H, B_2=g, f_1=q, \bar{f}_2=\bar{q}$, where $H$ is the lightest
Higgs boson and $q,\bar{q}$ are suitable quarks. The strong coupling
interpretation
\begin{equation}
a_1^{(2B^-)}=\alpha_s (E_2)
\end{equation}
yields $E_2=483.4(3)$ GeV =$m_H+2m_q$. However, experimental and
theoretical arguments \cite{pada} imply that the Higgs mass should
be in the region 100...200 GeV. Hence we only obtain a consistent
value for the Higgs mass if we assume that the quarks involved are
$t$ quarks. This is similar to the zero $a_1^{(2A^-)}$, where the
quarks involved were also heavy quarks. Generally, 
strings with antidiffusive couplings seem to describe
heavy rather than light particles.

In section 6.5 we will obtain quite a precise prediction of the
free top mass from the self energy of the chaotic 2B string,
namely $m_t=164.5(2)$ GeV, corresponding to a top pole mass of
174.4(3) GeV. With this value the zero $a_1^{(2B^-)}$ yields a
Higgs mass prediction of
\begin{equation}
m_H=E_2-2m_t=154.4(5) GeV.
\end{equation}
This is a very precise prediction, the statistical
error is very small. But of course the main source of uncertainty is a theoretical
uncertainty, namely
whether our Feynman web interpretation of the zero $a_1^{(2B^-)}$ is correct.
For example, assuming that a supersymmetric extension of the
standard model is correct, then the zero could also describe another scalar
particle.

Similar as for the $N=3$ strings,
the next larger zeros $a_2^{(2A^-)}$ and $a_2^{(2B^-)}$ should be considered as well.
What is interesting here is the fact
that near $a=\frac{1}{2}$ $(2A^-)$ and near $a=1$ $(2B^-)$ all
chaotic fluctuations cease to exist, and the fields $\Phi_n^i$
approach a stable homogeneous fixed point $0$ at
$a_2^{(2A^-)}=\frac{1}{2}$, respectively $a_2^{(2B^-)}=1$.
From a stochastic quantization
point of view this means that there are no vacuum fluctuations any
more, and a purely classical theory arises if these strings are
used for 2nd quantization.

The zero $a_2^{(2A^-)}=\frac{1}{2}$ could be
interpreted as a dimensionless gravitational coupling strength,
defined by $\alpha_G=\frac{1}{2} (E/m_{Pl})^2$, at the lightest
black hole mass scale $E=m_{Pl}$. Here $m_{Pl}$
denotes the Planck mass. The backward coupling form with
$a_2^{(2B^-)}=1$ would then describe two such black holes.
There are no black holes lighter than of order of the Planck mass
$m_{Pl}$, since these immediately evaporate by Hawking radiation.

If this interpretation is correct, then the smallest zeros of the
interaction energies $W(a)$ of the chaotic string theories fix the
strengths of all four interactions. Electroweak interactions are
fixed at the smallest fermionic mass scales $(N=3)$, strong
interactions at the smallest bosonic mass scales ($N=2$),
gravitational interactions at the smallest black hole mass scales
($N=2,s=-1$).

\section{Local minima of the self-energy $V(a)$}

\subsection{Self-interacting scalar field equations}

Let us now look at the other type of vacuum energy, the
self-energy $V$ of the chaotic strings. As mentioned at the end of
section 3, we expect that the self-energy is locally extremized
for physically observable states.


One subset of free standard model parameters (the coupling
strengths of the four interactions) was already fixed by the zeros
of the interaction energy $W$, but for another subset of free
parameters (essentially masses) the
self-energy $V$ may be the relevant quantity to look at. Consider
an {\em a priori} free standard model coupling $\alpha$ that
depends on such a free mass parameter. For example,
$\alpha$ may be a Yukawa coupling. Again we assume that this
standard model coupling $\alpha$ is a coupling $a$ in string space
at the same time. We may regard $\alpha =a$ as a homogeneous
self-interacting scalar field variable (similar to a dilaton field
in superstring theory) with a complicated effective potential
given by the expectation of the self energy $V(a)$ of the chaotic
string. The stochastically quantized field equation reads
\begin{equation}
\dot{a}=-\Gamma \frac{\partial V}{\partial a} +noise, \label{evo2}
\end{equation}
and if $\Gamma >0$ then local minima of the function $V(a)$ describe
possible stable stationary states of standard model couplings. Any
minimum of $V$ is thus a candidate for observed standard model
couplings. We will denote the local minima of $V(a)$ as $a_i'^{(Nb)}$.

Indeed a large number of local minima is observed that precisely coincide with known
standard model interaction strengths. The reader not interested
in all the details may directly proceed to section 6.6.

\subsection{The 3A string ---weak and strong interactions
of heavy fermion flavours}

Let us again start with the 3A string. Fig.~9 shows the
numerically determined function $V(a)$. First let us first look at
rather large values of the coupling $a$. The minima labelled as
\begin{eqnarray}
a_6'^{(3A)}&=&0.0953(1) \\ a_7'^{(3A)}&=&0.1677(5) \\
a_{8}'^{(3A)}&=&0.2327(5)
\end{eqnarray}
seem to coincide with strong couplings at the heavy quark mass
scales. A suitable Feynman web interpretation is given by $B_1$
$massless$, $B_2=g$, $f_1=q$, $\bar{f_2}=\bar{q}$, with $q=t,b,c$,
respectively. Formula (\ref{katharina}) with $N=3$ implies
\begin{equation}
a_i'^{(3A)}=\alpha_s(3m_q).
\end{equation}

The experimentally measured values of the free quark masses are
$m_c=1.26(3)$ GeV, $m_b=4.22(4)$ GeV, $m_t=164(5)$ GeV (these
numerical values are obtained if one performs the average over the
various results quoted in \cite{pada}).
With the scale parameters $\Lambda^{(n_f)}$ determined earlier in
section 5.5, the QCD formula in Appendix B yields the
corresponding strong couplings as
\begin{eqnarray}
 \alpha_s (3 m_t ) &=& 0.0952(3)\\
 \alpha_s (3 m_b ) &=& 0.1684(4)\\
 \alpha_s (3 m_c ) &=& 0.2323(20).
\end{eqnarray}
Apparently
there is good coincidence with the observed local minima of the
self energy of the 3A-string.

Further minima are observed that seem to describe weak interaction
strengths of right-handed fermions. Remember that in the standard
model the weak coupling constant is given by
\begin{equation}
\alpha_{weak} = \alpha_{el} \frac{(T_3-Q\sin^2 \theta_W)^2}{\sin^2 \theta_W
\cos^2\theta_W},
\end{equation}
where $Q$ is the electrical
charge of the particle ($Q=-1$ for electrons, $Q=2/3$
for $u$-like quarks, $Q=-1/3$ for $d$-like quarks),
and $T_3$ is the third component of the isospin ($T_3=0$ for right-handed
particles, $T_3=-\frac{1}{2}$ for $e_L$ and $d_L$, $T_3=+\frac{1}{2}$
for $\nu_L$ and $u_L$).
Consider right-handed fermions $f_R$. With $\sin^2 \theta_W
=0.2318$ and the running electric coupling $\alpha_{el} (E)$ taken at energy
scale $E=3 m_f$
we obtain
\begin{eqnarray}
\alpha_{weak}^{d_R} (3 m_d) &=&0.000246 \\
\alpha_{weak}^{c_R} (3 m_c) &=&0.001013 \\
\alpha_{weak}^{e_R} (3 m_e) &=&0.00220 .
\end{eqnarray}
On the other hand, we observe that $V(a)$ has minima at
\begin{eqnarray}
a_{1}'^{(3A)} &=& 0.000246(2) \\ a_{2}'^{(3A)} &=& 0.00102(1)
\\ a_{3}'^{(3A)} &=& 0.00220(1)
\end{eqnarray}
(see Fig.~9b. $a_1'$ and $a_3'$ are actually small local minima on top of the hill).
There is good agreement with the weak coupling
constants of $f_R=u_R,c_R,e_R$, respectively. A suitable Feynman
web interpretation would be $B_1$ {\em massless}, $B_2=Z^0, f_1=f_R, \bar{f_2}=\bar{f_R}$.
Basically,
the minima yield statements on the charges of the
particles involved. Generally,
it is reasonable to assume that
the above interaction states of $d,c,$ and $e$ are not pure mass eigenstates but
that small components of other flavours are mixed in as well.
For example, $c$ could also have a small $t$ component, thus slightly
increasing the relevant energy scale.

If $a_1'^{(3A)},a_2'^{(3A)},a_3'^{(3A)}$ decribe right-handed fermions
of type $d_R,u_R,e_R$ (mixed with higher flavors) then it is natural
to assume that the next minimum $a_4'^{(3A)}=0.00965(1)$ decribes
a right-handed neutrino $\nu_R$. This could be
a very heavy right-handed Majorana neutrino interacting gravitationally
with
$\alpha_G=\frac{1}{2}(m_{\nu_R}/m_{Pl})^2=a_4'^{(3A)}$. From this
one obtains
the mass $m_{\nu_R}=1.696(1)\cdot 10^{18}$ GeV. We will come
back to neutrinos in section 6.7. 

The remaining
minimum $a_5'^{(3A)}=0.02145(3)$ inbetween weak
minima $a_1'^{(3A)},a_2'^{(3A)},a_3'^{(3A)}$
and strong minima $a_6'^{(3A)},a_7'^{(3A)},a_8'^{(3A)}$
could decribe a unified coupling
at the GUT scale. In fact, evolving the strong
coupling $\alpha_s(3m_t)=a_6'^{(3A)}$
to higher energies using the usual (non-supersymmetric)
formulas \cite{xxx}, one obtains $\alpha_s(3m_Q)=a_5'^{(3A)}$ for 
$m_Q\approx 1.74\cdot 10^{16}
$ GeV, which is the GUT scale. At $E=m_Q$ the strong coupling
$\alpha_s(E)$ and the
weak coupling $\alpha_2(E)$ are observed to merge into the common value
$\alpha_s=\alpha_2=0.02192$.


\subsection{The 3B string --- further mixed states of heavy fermion
flavours}

$V(a)$ for the 3B  string is plotted in Fig.~10. This string is of
the backward coupling form, and our general arguments suggested
that the backward coupling form describes mixed states of
two particle species. We observe (among others) the following minima of
the vacuum energy.
\begin{eqnarray}
a_6'^{(3B)}=0.1027(1)  \label{mix1} \\ a_8'^{(3B)}=0.2916(5)
\label{mix2}.
\end{eqnarray}
If we choose the same numerical
values of the quark masses and the parameters $\Lambda^{(n_f)}$ as before,
we obtain from the QCD-formula (\ref{runas})
\begin{eqnarray}
\alpha_s \left(\frac{3}{2}(m_t+m_b)\right) &=& 0.1027(4) \\
\alpha_s \left(\frac{3}{2}(m_c+m_s)\right) &=& 0.2914(35) .
\end{eqnarray}
For the $s$-quark mass we used the value $m_s=(160 \pm 9)$ MeV
\cite{nari}. Again there is good coincidence between the above
standard model couplings and the observed string couplings. The
zeros $a_6'^{(3B)}$ and $a_{8}'^{(3B)}$ illustrate that up- and
down quark flavours can mix flavour-independently for strong
interaction.

At lower coupling strengths, the minimum $a_3'^{(3A)}=0.00220$ of the 3A
string slightly changes to $a_3'^{(3B)}=0.00223$ for the 3B string.
This numerically coincides with the weak coupling
$\alpha_{weak}^{e_R} (\frac{3}{2}(m_e+m_\mu))=0.00223$ of a state
where $e$ and $\mu$ mix with equal weights. If this interpretation
is correct, then the minimum $a_3'^{(3B)}$ can be used to estimate
the $\mu$ mass as being of the order of magnitude 100 $MeV$.
Similarly, the minimum $a_4'^{(3A)}$ describing the
heavy right-handed Majorana neutrino changes by a small amount
to $a_4'^{(3B)}=0.00972(1)$, which could be due to neutrino
flavor mixing.

Also the minima $a_1'^{(3A)}$ and $a_2'^{(3A)}$ of the 3A string change by a
very small positive amounts to $a_1'^{(3B)}$, $a_2'^{(3B)}$ for the
3B string. It is difficult to numerically estimate this change, it
is of the order of the precision by which we can determine the
minima. It could again mean that small amounts of heavier flavours
are mixing with $d,c$. For example, $d$ might get a small $s$
component and $c$ a small $t$ component. In total, one may
conjecture that the four minima $a_1'^{(3A)},a_2'^{(3A)},a_1'^{(3B)},
a_2'^{(3B)}$
contain the information on the four mixing angles of the
Kobayashi-Maskawa matrix.

At larger couplings $a$, the $N=3$ strings exhibit lots of further local
minima. There is evidence that many of these can be associated with
baryonic and mesonic states.

\subsection{The 2A string ---
further bosons}
Let us now turn to $N=2$ strings.
The self-energy of the 2A string theory is
shown in Fig.~11. A very strongly pronounced minimum is
\begin{equation}
a_3'^{(2A)}=0.1848(1)
\end{equation}
It coincides with the strong coupling $\alpha_s$
at twice the $b$-quark mass. The Feynman web
interpretation of this chaotic string state could be
$B_1=B_2=g$, $f_1=b,\bar{f_2}=\bar{b}$. Since $N=2$ the
relevant energy scale is $E=(N/2)(m_g+2m_b)=2m_b$.
Numerically, we get
from eq.~(\ref{runas})
for $m_b=4.23$ GeV
\begin{equation}
\alpha_s (2m_b)= 0.1848,
\end{equation}
which coincides precisely with the minimum $a_3'^{(2A)}$. In fact,
the minimum is very sharp, moreover, it is in the region where the
scale parameter $\Lambda^{(5)}$ is of relevance, which we know
with high precision from section 5.5. Hence we may go the other
way round and use $a_3'^{(2A)}$ to predict $m_b$ with high
precision. The value $\Lambda^{(5)}=0.20608(14)$ GeV obtained
earlier translates $a_3'^{(2A)}=0.1848(1) =\alpha_s(2m_b)$ to 
\begin{equation}
m_b=4.23(1) GeV.
\end{equation}
This is in good agreement with estimates of the free $b$-mass using
current algebra techniques \cite{nari}.

Another minimum of interest is
\begin{equation}
a_2'^{(2A)}= 0.03369(2)
\end{equation}
This seems to coincide with the weak coupling constant $\alpha_2
\approx 1/30$ at the $Z^0$ mass scale. Our Feynman web
interpretation is $B_1=Z^0$, $B_2=W_\mu^0$, $f_1=b, f_2=\bar{b}$,
which implies
\begin{equation}
a_2'^{(2A)}=\alpha_2 (m_{Z^0}+2m_b).
\end{equation}
We can estimate the value of the weak coupling $\alpha_2$ at this
energy scale using the well-known formulas for the running
electroweak couplings \cite{xxx}. At the $Z^0$ mass scale
$m_{Z}=91.188$ GeV the QED formula (\ref{runael}) yields
\begin{equation}
\alpha_{el}(m_Z)=0.0078096
\end{equation}
The running electroweak mixing angle defined as
$\hat{s}^2(E):=\alpha_{el} (E)/ \alpha_2 (E)$ has been shown to
very closely coincide with the effective mixing angle $\sin^2
\theta_{eff}^{lept}$ at the energy $E=m_{Z}$. One has \cite{pada}
\begin{equation}
\hat{s}^2 (m_{Z}) = \sin^2 \theta_{eff}^{lept} -0.00029
\end{equation}
Hence our result of $\sin^2 \theta_{eff}^{lept}=0.23177(7)$
derived in section 5.4 yields
\begin{equation}
\hat{s}^2(m_{Z})=0.23148(7),
\end{equation}
which implies
\begin{equation}
\alpha_2 (m_{Z}) = \frac{\alpha_{el}
(m_{Z})}{\hat{s}^2}=0.03374(1).
\end{equation}
If further transferring the running $\alpha_2$ to the slightly
larger energy scale $m_Z+2m_b$ one obtains the numerical value
\begin{equation}
\alpha_2(m_Z+2m_b)=0.03369(1)
\end{equation}
which precisely coincides with the observed string coupling
${a_2}'^{(2A)}$ with a precision of 4 digits.

Finally, there is also a rather weakly pronounced minimum
$a_1'^{(2A)}=0.00755(3)$ which could be interpreted as
$\alpha_{el}(2m_\tau)$. In the Feynman web interpretation
$B_1=B_2=\gamma$, $f_1=\tau$, $\bar{f_2}=\bar{\tau}$. This minimum could
be used to estimate that the order of magnitude of the $\tau$ mass is
about 2 GeV.

Summarizing, we see that whereas the interaction energy of the 2A
string provided evidence for the charged gauge bosons $W^\pm$, the
self energy of the 2A string provides evidence for the uncharged
gauge bosons $g$, $Z^0$ and $\gamma$.

But clearly there are much more local minima of the self energy
than the above three minima that we analyzed in detail. In fact,
local minima of $V(a)$ accumulate near $a=0$ for all
six chaotic string theories. It is unlikely that
for all those minima simple standard model interpretations can be
found. But we can always associate an observed minimum $a_i'$ with
a particle of mass $m_i$ that interacts by gravitational
interaction with $\alpha_G=\frac{1}{2}(m_i/m_{Pl})^2=a_i'$. In
this sense the many minima observed could simply mean that there
are lots of particles that purely interact gravitationally but do
not take part in standard model interactions. These could then be
associated with dark matter. Chaotic strings seem to predict a
very broad (though discrete) spectrum of dark matter particles,
with masses covering all orders of magnitude.

\subsection{The 2B string --- Yukawa interaction
of the top quark}

Let us return to minima for which a standard model interpretation
can be found. The self energy of the 2B string is shown in
Fig.~12.

A very interesting
minimum observed is
\begin{equation}
a_2'^{(2B)}=0.03440(2).
\end{equation}
This string coupling can be interpreted in terms of Yukawa interaction of the top quark.
Our Feynman web
interpretation is $B_1=B_2=H$, $f_1=t, f_2=\bar{t}$.


The Yukawa coupling
of any fermion $f$ is proportional to the square of its mass. It is given by
\begin{equation}
\alpha_{Yu}^f= \frac{1}{4} \alpha_2 \cdot \left( \frac{m_f}{m_W} \right)^2.
\end{equation}
Still we have to decide on the energy scale $E$ for the running
$\alpha_2 (E)$. Our usual rules for the Feynman web interpretation
$B_1=B_2=H$, $f_1=t$, $f_2=\bar{t}$ imply that the energy is given
by $E=m_H+2m_t$. Our interpretation of the minimum
$a_2'^{(2B)}$ is thus
\begin{equation}
a_2'^{(2B)}= \alpha_{Yu}^t (m_H+2m_t)=\frac{1}{4} \alpha_2
(m_H+2m_t) \left( \frac{m_t}{m_W} \right)^2.
\end{equation}
Accepting this interpretation the formula allows for a very
precise prediction of the top-mass $m_t$. From our earlier
consideration we know that $\alpha_2 (m_Z+2m_b)=0.03369(1)$.
Transferring the running $\alpha_2$ to the higher energy scale
$E=m_H+2m_t\approx 2m_W+2m_t$ we obtain
\begin{equation}
\alpha_2 (E)=0.03284(1)
\end{equation}
(the usual (non-supersymmetric) formula for $\alpha_2(E)$ is used \cite{xxx}).
Solving for $m_t$ we get
\begin{equation}
\frac{m_t}{m_W} = 2 \sqrt{\frac{a_2'^{(2B)}}{\alpha_2 (E)}} =
2.047(1)
\end{equation}
From the value $m_W=80.35(6)$ GeV
we thus get
\begin{equation}
m_t=164.5(2) GeV,
\end{equation}
The corresponding pole mass $M_t$, of relevance for experiments,
can be quite precisely determined by a formula in
Degrassi et al. \cite{degra}, which is based on $\Lambda^{(5)}$ and
avoids $\Lambda^{(6)}$. We get
\begin{equation}
M_t=174.4(3) GeV
\end{equation} The error only takes into account the
precision by which we can determine $a_2'^{(2B)}$, in addition
there is the theoretical uncertainty whether our Feynman web
interpretation of the minimum is correct. Nevertheless, our prediction
coincides with the experimentally measured value
$M_t=(174.3 \pm 5.1)$ GeV \cite{pada} and is in fact more precise.

Let us estimate the largest possible systematic error of our prediction of the
top-mass. The value of the Higgs mass is uncertain, for example it
could turn out to be of the order 100 GeV rather than our
conjectured value of 154 GeV. Also the error in $\alpha_2 (m_Z)$ could
be slightly underestimated. But all this would only mean that our
top-mass prediction changes by less than $1$ GeV.

\subsection{Yukawa and gravitational interactions of all quarks
and leptons}

Clearly not only the top-quark, but also the other heavy fermions are able to
exhibit
Yukawa interaction. The corresponding couplings are much smaller, since they
all go with the mass squared of the particles involved. Nevertheless,
they are indeed observed as suitable
minima of the self-energy of the 2A/2B string.

Generally, for all chaotic strings scaling behavior sets in if the
coupling $a$ approaches 0,
and in this limit there is no difference between the forward and
backward coupling form. One numerically observes for $a\to 0$
\begin{equation}
V(a)-V(0)=f^{(N)}(\ln a) \cdot a^{\frac{1}{2}},
\end{equation}
where $f^{(N)}(\ln a)$ is a periodic function of $\ln a$ with
period $\ln N^2$. Hence in a double logarithmic plot of
$|V(a)-V(0)|$ versus $a$ one observes a straight line that is
modulated by oscillating behaviour. From the periodicity it
follows that if there is some local minimum at $a_i$ then there is
also a minimum at $a_i/N^{2n}$ for arbitrary $n$. Thus all local
minima in the scaling region are only determined modulo $N^2$.

$|log_9 |V(a)-V(0)||$ is plotted for the 3A/B string in Fig.~13, and
$|log_4 |V(a)-V(0)||$ for the $2A/B$ string in Fig.~14. 
One has $V(0)=\frac{3}{8}$ for $N=3$ and $V(0)=0$ for $N=2$.
Whereas the 3A/B string has only 2 minima
per period (essentially describing the charge ratios of $d,u,e$,
see section 6.2), the 2A/B string exhibits a much richer structure.
For simplicity, let us denote the local minima of the 2A/B string in
the scaling region as $b_i$. Within one period
of length $ln 4$ , one observes 11 different minima (Fig.~15).
These we have numerically determined in the region
$[0.000143,0.000572]$ as
\begin{eqnarray*}
b_1&=&0.000199(1) \\ b_2&=&0.000263(2) \\
b_3&=&0.000291(1) \\ b_4&=&0.000306(1) \\
b_5&=&0.000345(1) \\ b_6&=&0.000368(3) \\
b_7&=&0.000399(1) \\ b_8&=&0.000469(1) \\
b_9&=&0.000482(1) \\ b_{10}&=&0.000525(2) \\
b_{11}&=&0.000558(1)
\end{eqnarray*}
(only $b_i \; mod \; 4$ is relevant).
The remarkable fact is that these local minima of the self energy
can be associated with Yukawa and gravitational couplings of all
quark and lepton flavours modulo 4. Heavy particles turn
out to minimize Yukawa couplings\footnote{Note
that we do observe the Yukawa couplings of the ordinary
standard model and not those of a supersymmetric extension.},
light particles gravitational couplings. Leptons are found
in the left part of Fig.~15, quarks in the right part (modulo 4).

Let us start with the heavy fermions $t,b,\tau ,c$. The Yukawa
coupling of the $t$ quark was already described by the minimum
$a_2'^{(2B)}=0.03440=:\tilde{b_5}$ outside the scaling region. This minimum
can be formally regarded as evolving out of $b_5$ in the scaling
region.
The
minima $b_2,b_6,b_{10}$ turn out to coincide
with Yukawa couplings modulo 4 of $\tau ,b ,c$, respectively. We
observe
\begin{equation}
b_i=\alpha_{Yu} =\frac{1}{4} \alpha_2 (m_H+2m_f) \left(
\frac{m_f}{m_W} \right)^2 \cdot 4^n,
\end{equation}
where $f$ denotes the fermion under consideration.
Solving for $m_f$ we get a prediction of the heavy fermion mass
modulo 2.
\begin{equation}
m_f=\sqrt{\frac{b_i}{\alpha_2(m_H+2m_f)}}m_W 2^{-n+1}
\end{equation}
One obtains from the observed minima the following results.

\vspace{0.5cm}

\begin{tabular}{l|l|l|l|l|l}
$f$ & $\alpha_2(m_H+2m_f)$ & $i$ & $b_i \cdot 10^4$ & $n$ & $m_f$ [GeV]
\\ \hline $t$ & 0.03284 & (5) & 344.0(2) & 0 & 164.5(2) \\ $b$ & 0.03340 &
6 & 3.68(3) & 2 & 4.22(2) \\ $\tau$ & 0.03342 & 2 & 2.63(2) & 3 &
1.782(7)\\ $c$ & 0.03342 & 10 &5.25(2) & 4 & 1.259(4)
\end{tabular}

\vspace{0.5cm}

Note the small error in the mass predictions, which is due to the
fact that Yukawa couplings are very sensitive to the fermion
masses. On the other hand, the results are almost independent of
the Higgs mass, respectively the precise value of $\alpha_2(m_Z)$.
Of course, the integer $n$, which fixes the order of magnitude of
the predicted fermion masses, is not directly known from the
scaling region of the $N=2$ string. However, the information on
this integer is already given by other minima of other strings for
larger couplings. For example, the order of magnitude of the $b$
mass follows from the minimum $a_{7}'^{(3A)}$ or $a_3'^{(2A)}$,
that of the $c$ mass from the minimum $a_{8}'^{(3A)}$ or
$a_2'^{(3A)}$, and that of the $\tau$ mass from the minimum
$a_1'^{(2A)}$. Hence $n$ is known from other string states. We may
call $2^n$ a `winding number' of the chaotic string.

Now let us proceed to the lighter fermions, and to the
interpretation of the remaining minima $b_i$. Remarkably, for the
light fermions where the mass is known with rather high
precision\footnote{Averaging the various results on light quark
masses in the particle data listing \cite{pada}, one obtains the
values $m_u=4.7(9)$ MeV, $m_d=9.1(6)$ MeV, $m_s=167(7)$ MeV at the
proton mass renormalization scale.} we observe that the self
energy has local minima for couplings that coincide with
gravitational couplings modulo 4. We observe for $i=1,4,7,8,9$
\begin{equation}
b_i =\alpha_G = \frac{1}{2} \left( \frac{m_f}{m_{Pl}} \right)^2
\cdot 4^n,
\end{equation}
or, solving for $m_f$, we obtain the prediction
\begin{equation}
m_f=\sqrt{2b_i}\; m_{Pl}\; 2^{-n}
\end{equation}
of the light fermion masses. The result is shown in the following
table.

\vspace{0.5cm}

\begin{tabular}{l|l|l|l|l} $f$ & $i$ & $b_i \cdot 10^4$ & $n$ & $m_f$ [MeV]
\\ \hline

$\mu$ &1& 1.99(1) & 61 & 105.6(3) \\ e & 4& 3.06(1) & 69 &
0.5117(8) \\
 $d$ &7 & 3.99(1) & 65 &
9.35(1) \\  $u$ & 8 & 4.69(1) & 66 & 5.07(1) \\ $s$ & 9 & 4.82(1)
& 61 & 164.4(2) \\

\end{tabular}

\vspace{0.5cm}

Again the winding number $2^n$ determining the order of magnitude
of the mass is not known from the scaling region but already fixed by other string
states. For example, the zeros $a_2^{(3A)}, a_1^{(3A)}, a_1^{(3B)}$
of the interaction energy $W(a)$ yield
the order of magnitude of the masses of $e,d,u$, respectively.
The order of magnitude of the
$\mu$ mass follows from $a_3'^{(3B)}$,
and that of the $s$ mass from
$a_{8}'^{(3B)}$.



Generally, it is a very fortunate effect that scaling behaviour of
the vacuum energy sets in for small $a$. For example, we would
never be able to get any information on the gravitational coupling
of an electron $\alpha_G(m_e)=\frac{1}{2}(m_e/m_{Pl})^2= 8.76
\cdot 10^{-45}$ in a direct simulation with $a=\alpha_G$. However,
we can easily iterate the coupled map lattice with the much larger coupling
$a=\alpha_G(m_e) \cdot 4^{69}=0.000305$ and then conclude onto
$\alpha_G(m_e)$ modulo 4 via the scaling argument.

It is straightforward to conjecture that the remaining
minima can be associated with massive neutrino states.
This will be worked out in detail in the next section.

\subsection{Neutrino mass prediction}
It is desirable to find an interpretation of all minima that
possesses the highest symmetry standards. It is in fact possible
to provide a fully symmetric attribution of the 11 minima to the
12 known fermions if one assumes that the minimum $b_1$ is
degenerated, i.e.\ that it describes two different particles with
the same mass modulo 2. This scheme of largest symmetry is
described by the following table.

\vspace{0.5cm}

\begin{tabular}{c|c|c|c|c|c|c|c|c|c|c|c}
$b_0$ & $b_1^{(1)}$ & $b_1^{(2)}$ &
$b_2$&$b_3$&$b_4$&$b_5$&$b_6$&$b_7$&$b_8$&$b_9$&$b_{10}$ \\ \hline
$\nu_2$ &$\mu$ & $\nu_3$&$\tau$ & $\nu_1$ &
$e$&$(t)$&$b$&$d$&$u$&$s$&$c$
\end{tabular}

\vspace{0.5cm} Since all minima are only defined modulo 4, we have
identified $b_0 := b_{11}/4$. The $t$ quark is put into
parenthesis since $m_t$ is so large that the corresponding Yukawa
coupling falls out of the scaling region. Within the above scheme
all {\em up} and {\em down} members of the same family are
described by neighbored minima, and the {\em up} member always has
a larger self energy than the {\em down} member (see Fig.~15). All
leptons are grouped together (in family index order 2,3,1) and all
quarks as well (in family index order 3,1,2). The fact that for
the $N=2$ string one minimum is degenerated is completely
analogous and symmetric to the $N=3$ string, where also one
minimum was degenerated, describing the fact that electrons and
$d$-quarks have the same charge modulo 3 (see section 6.2).

The above scheme associates the three neutrino states
$\nu_1,\nu_2,\nu_3$ with the minima $b_3,b_0,b_1^{(2)}$,
respectively, and hence these minima fix the neutrino masses
modulo 2 as
\begin{eqnarray}
m_{\nu_1}&=&\sqrt{2b_3}\; m_{Pl}\; 2^{-n_1} \label{loft1} \\
m_{\nu_2}&=&\sqrt{2b_0}\; m_{Pl}\; 2^{-n_2} \label{loft2}
\\ m_{\nu_3}&=&\sqrt{2b_1^{(2)}}\; m_{Pl}\; 2^{-n_3}, \label{loft3}
\end{eqnarray}
or equivalently
\begin{eqnarray}
m_{\nu_1}&=&0.952(1)eV \;\; mod \; 2 \label{mod1}
\\ m_{\nu_2}&=&1.318(1)eV \;\; mod \; 2 \label{mod2}\\
m_{\nu_3}&=&1.574(4)eV \;\; mod \; 2 \label{mod3}.
\end{eqnarray}
Since the relevant string coupling constant is the gravitational
coupling $\alpha_G=\frac{1}{2}(m_\nu/m_{Pl})^2$ one expects that
these values represent the masses of the mass eigenstates
$\nu_1,\nu_2,\nu_3$ (rather than those of the weak eigenstates
$\nu_e,\nu_\mu,\nu_\tau$).

To obtain concrete numerical values we have to decide on the
relevant powers of 2. Let us here be guided by symmetry
considerations with the other fermions. Define the open intervals
$I_1$ and $I_2$ as
\begin{equation}
I_1=(2^4,2^5) \;\;\;\;\;\; I_2=(2^7,2^8)
\end{equation}
We observe
\begin{eqnarray}
\frac{m_t}{m_c}=130.7 \in I_2 \\ \frac{m_c}{m_u}=248.3 \in I_2 \\
\frac{m_b}{m_s}=25.67 \in I_1 \\ \frac{m_s}{m_d}=17.58 \in I_1 \\
\frac{m_\tau}{m_\mu}=16.83 \in I_1 \\ \frac{m_\mu}{m_e}=206.7 \in
I_2,
\end{eqnarray}
i.e. all $up$-type quark mass ratios are in $I_2$, all $down$-type
quark mass ratios in $I_1$, 3rd/2nd family lepton mass ratios are
in $I_1$, 2nd/1st family lepton mass ratios in $I_2$. The most
reasonable assumption giving full symmetry to the problem is to
assume that neutrinos follow this general pattern as well, i.e.
\begin{eqnarray}
\frac{m_3}{m_2} \in I_1 \label{i1} \\
\frac{m_2}{m_1} \in I_2 \label{i2}
\end{eqnarray}
So far the most stringent experimental evidence for neutrino masses comes from
atmospheric neutrinos, providing evidence for
\begin{equation}
\Delta m_a^2 \approx 3\cdot 10^{-3} eV^2 \label{expe}
\end{equation}
and maximal mixing \cite{pada}. If there is a hierarchy of neutrino
masses $m_{\nu_3}>>m_{\nu_2}>>m_{\nu_1}$ then the experimental
result can be interpreted as $m_{\nu_3} \approx \sqrt{\Delta
m_a^2}\sim 0.055 eV$. If this experimental estimate is correct
with a precision of a factor 2, then eq.\ (\ref{loft3}) together
with (\ref{expe}) implies $n_3=92$, and eqs.~(\ref{loft2}),
(\ref{loft1}) together with (\ref{i1}), (\ref{i2}) imply
$n_2=96,n_1=104$. This means that the chaotic string spectrum
yields the very precise predictions
\begin{eqnarray}
m_{\nu_1}&=&1.452(3)\cdot 10^{-5}eV \label{m1} \\
m_{\nu_2}&=&2.574(3)\cdot 10^{-3}eV \label{m2}\\
m_{\nu_3}&=&4.92(1)\cdot 10^{-2}eV. \label{m3}
\end{eqnarray}

We may also try to avoid the experimental input (\ref{expe}) and
use a purely theoretical argument to fix the relevant energy
scale. As described in section 6.2, the minimum $a_4'^{(3A)}$ can
be theoretically interpreted as indicating the existence of a very
heavy right-handed neutrino $\nu_R$ of mass
$m_{\nu_R}=1.696(1)\cdot 10^{18}$ GeV. The seesaw mechanism 
\cite{gellmann,falcone}
provides an estimate of the light (left-handed) neutrino masses
from the heavy right-handed ones via the equation
\begin{equation}
m_{\nu_L}\approx\frac{m_q^2}{m_{\nu_R}}, \label{seesaw}
\end{equation}
where $m_q$ is of the order of magnitude
of a typical quark mass. Choosing $m_q=m_t$
eq.~(\ref{seesaw}) yields $m_{\nu_L}\approx 1.59\cdot 10^{-5}$ eV.
This is very close to the value of $m_{\nu_1}$ in eq.~(\ref{m1}),
thus suggesting that $\nu_{L}$ is an electron neutrino and that
there is almost no mixing of $\nu_1$ with $\nu_2$ (or $\nu_3$) to
form $\nu_e$, i.e.\ $\nu_e \approx \nu_1$. In this way one again
ends up with $n_1=104$ and eqs.~(\ref{m1})--(\ref{m3}) are
obtained by purely theoretical arguments. The predicted value for
$m_{\nu_2}$ is consistent with present experimental measurements,
i.e.\ the low-mixing angle (LMA) solution of the solar neutrino
problem. The experiments provide evidence for 
$\Delta m_s^2\approx 5.4\cdot 10^{-6}\;{eV}^2$ \cite{pada}, which could be
interpreted as $m_{\nu_2}\approx 2.3\cdot 10^{-3}\;eV$, in very good
agreement with eq.~(\ref{m2}).

\subsection{The $2A^-$ and $2B^-$ strings --- mass ratio of heavy bosons}

The fermion mass ratios were determined by the minima $b_i$ of the
2A/2B string, according to the relation
\begin{equation}
\frac{b_{i_1}}{b_{i_2}}=\left( \frac{m_{f_1}}{m_{f_2}} \right)^2 mod \; 4
\end{equation}
The quadratic dependence comes from the quadradic energy dependence
of gravitational (or Yukawa) couplings.

It now seems plausible that also boson mass ratios could be fixed
in a similar way, i.e.\ by the self energy of some suitable string
theories in the limit of very small $a$. The most plausible
remaining candidates are the 2A$^-$/2B$^-$ strings. The self
energy of these strings also exhibits periodic scaling behaviour with
period $\ln \; 4$
for $a\to 0$, but for the anti-diffusive coupling form only two
minima per period are observed, which we have determined
in the region [0.000143,0.000572] as $b_1^-=0.000335(1)$
and $b_2^-=0.000361(2)$ (Fig.~16). In analogy to the fermionic case
we may assume that the bosonic mass ratios are given by
\begin{equation}
\frac{b_1^-}{b_2^-}=\left( \frac{m_{B_1}}{m_{B_2}} \right)^2 mod
\; 4.
\end{equation}
Suppose one of the minima describes the $W$-boson, of mass
$80.35(6)$ GeV, and the other one the Higgs boson. Then, depending
on which minimum is identified with which particle as well as the
unknown power of 4, the above equation allows for the following
masses of the Higgs boson: 77.4, 83.4, 154.8, 166.8, 309.6, 333.6
GeV. But experimental and theoretical bounds \cite{pada} imply 95
GeV $<m_H<$ 190 GeV, hence only the values 154.8(7) or 166.8(7)
GeV survive. It is remarkable that (within the error bars) the
value 154.8(7) GeV coincides with the value 154.4(4) GeV that was
predicted independently in section 5.7. This value implies that
the minimum $b_1^-$ is attributed to $W$ and $b_2^-$ to $H$.
Numerically, one observes the relation
\begin{equation}
b_i^-=\frac{1}{2}a_4'^{(3A)} \left( \frac{m_{Pl}}{m_{B_i}} \right)^2 
\; mod \; 4= \left(\frac{m_{\nu_R}}{m_{B_i}}\right)^2 \; mod \; 4,
\end{equation}
which suggests that what is really fixed by the $2A^-/2B^-$ string is the ratio $m_{\nu_R}/m_{B_i}$
between fermion and boson masses. Note that
the coupling $b_i^-$ is proportional to 
an {\em inverse} gravitational coupling. This reminds us
of the concept of duality in superstring theories, where couplings are
replaced by inverse couplings.

\section{Summary and Outlook}

Instead of considering the standard model alone and putting in
about 25 free parameters by hand, in this paper we have suggested
to postulate the existence of chaotic strings underlying the noise
of the Parisi-Wu approach of stochastic quantization on a very
small scale.
The chaotic string dynamics can be
physically interpreted as a 1-dimensional strongly fluctuating
dynamics of vacuum fluctuations. It generates effective potentials
which distinguish the observed standard model couplings from
arbitrary ones. The dynamics may either have determined
the standard model parameters at a very early stage of the
universe (e.g.\ in a pre-big bang scenario) or it may still evolve
today and stabilize the observed values of the parameters.


Whereas for standard model fields, as well as for superstrings
after compactification, continuous gauge symmetries such as
$U(1)$, $SU(2)$ or $SU(3)$ are relevant, for the chaotic strings
a discrete $Z_2$ symmetry is relevant.
Whereas standard model fields or ordinary strings usually evolve
in a regular way, the chaotic strings obtained for $N>1$ evolve in
a deterministic chaotic way\footnote{For $N=1$ there is no chaotic
behavior and one just obtains a discretized heat equation.}. They
are strongly self-interacting and correspond to a Bernoulli shift
of information for vanishing spatial coupling $a$. The constraint
conditions on the vacuum energy (or the analogues of the Einstein
and scalar field equations) fix certain equilibrium metrics in
string space, which determine the strength of the Laplacian
coupling. We have provided extensive numerical evidence that these
equilibrium metrics reproduce the free standard model parameters
with very high precision. Essentially coupling constants are fixed
by the interaction energy $W(a)$, and masses, mass mixing
angles and charges by the self energy $V(a)$. This is summarized in Fig.~17.

The simplest physical interpretation is to regard the chaotic
string dynamics as a dynamics of vacuum fluctuations, which is 
present everywhere but which is unobservable due to the
uncertainty relation. Only expectations of the dynamics can be
measured, in terms of the fundamental constants of nature.
The dynamics may have fixed the free parameters already at a
very early (pre-big bang) stage of the universe.

%

On the other hand, one may also wish to embedd
chaotic strings into more general
theories, for example M-theory.
Several different approaches seem possible. One might
try to attribute to each of the 6 components
that make up M-theory in moduli space one of the 6
chaotic string theories, used for second quantization. A possible
picture is plotted in Fig.~18. Many other scenarios seem possible,
and it is clear that at the present stage any diagram of the
type of Fig.~18 is merely speculation. What, however,
is clear is that an ordinary string winding around a compactified space has a
discrete momentum spectrum, and the string field variable
$X^\mu (\sigma ,\tau)$ is a kind of position variable
taking on continuous values. On the contrary, a chaotic
string has a discrete position spectrum $i$ and the field variable
$\Phi_n^i$ is a kind of momentum variable taking on continuous values.
Hence in that sense the role of position and momentum is
exchanged.

An interesting possibility is that after compactification each of the 6 chaotic
strings might wind around one of the 6 compactified dimensions,
respecting the relevant (unknown) topological structure
of the compactified space. Note that if
chaotic strings live in the compactified space then they do not
`disturb' our usual understanding of 4-dimensional space-time
physics. Rather, they yield a very relevant amendment. Each
inverse string coupling constant $a^{-1}$ can be regarded as an
element of the metric in the compactified space, and the analogues
of the Einstein- and scalar field equations make the observed
standard model parameters evolve to the minima of the effective
potentials. This would give physical meaning to the compactified
space, in the sense that at each ordinary space-time point the
standard model parameters are known and in fact stabilized, due to the chaotic dynamics
in the compactified space.

Generally, the chaotic strings exhibit symmetry under the
replacement $T_N \to -T_N$, which could be formally associated
with a supersymmetry transformation. However, when introducing the
evolution equations (\ref{evo}) and (\ref{evo2}) of the couplings
one has to decide on the sign of the constant $\Gamma$. This choice
effectively breaks supersymmetry.
Generally, it seems reasonable to assume that with such a
choice of sign the expectation of
the vacuum energy of the chaotic noise strings effectively breaks
supersymmetry and singles out the physically relevant string vacua
of superstring theory. Supersymmetric partners of
ordinary particles might
be formally described by maxima rather than minima of the effective potentials
of the chaotic strings --- but these string states are unstable states
in fictitious time.
The instability might indicate that supersymmetric partners,
though formally there to cancel divergences in the Feynman diagrams as well as
unwanted vacuum energy,
may turn out to be unobservable in our world.

\section{Appendix A: large coupling limit of a $\phi^4$-theory}
Consider a
$\phi^4$-theory in 1 dimension that is stochastically
quantized. The field equation is
\begin{equation}
\frac{\partial}{\partial t} \phi (x,t) = \left(
\frac{\partial^2}{\partial x^2} -\mu^2 \right) \phi (x,t) -\lambda
\phi^3 (x,t) + noise . \label{pro26}
\end{equation}
Discretizing space $x$ with lattice constant $\delta$ and fictitious
time $t$ with lattice constant $\tau$ one obtains
\begin{equation}
\frac{\Phi_{n+1}^i-\Phi_n^i}{\tau} = \frac{\Phi_n^{i+1} \label{3a}
-2\Phi_n^i+\Phi_n^{i-1}}{\delta^2}-\mu^2\Phi_n^i-\lambda {\Phi_n^i}^3+
noise.
\end{equation}
This can be written as
\begin{equation}
\Phi_{n+1}^i= (1 -2\frac{\tau}{\delta^2}-\mu^2 \tau )\Phi_n^i -\lambda
\tau \Phi_n^{i^3} +\frac{\tau}{\delta^2} (\Phi_n^{i+1} +\Phi_n^{i-1})
+\tau \cdot noise \label{q-2}
\end{equation}
Now let $\tau \to 0$, $\delta \to 0$, $-\mu^2 \to \infty$, $\lambda \to
\infty$ such that
\begin{eqnarray}
\frac{\tau}{\delta^2}&=:&\frac{a}{2}\;\; \mbox{   finite} \\ \mu^2 \tau
&=:& \mu^2_{ren} \;\;\mbox{   finite} \\ \lambda \tau &=:&
\lambda_{ren}\;\; \mbox{   finite.}
\end{eqnarray}
Eq.~(\ref{q-2}) then reduces to
a coupled map lattice
of the form
\begin{equation}
\Phi_{n+1}^i =(1-a) T(\Phi_n^i )+\frac{a}{2}
(\Phi_n^{i+1}+\Phi_n^{i-1}) . \label{dyna}
\end{equation}
The local map $T$ is given by
\begin{equation}
T(\Phi )= \left( 1 -\frac{\mu^2_{ren}}{1-a} \right) \Phi
-\frac{\lambda_{ren}}{1-a}\Phi^3. \label{map}
\end{equation}
The noise term $\tau \cdot noise$ actually vanishes in the limit
$\tau \to 0$ that is considered here. A local map with
strongest possible chaotic behaviour is obtained for the choice
\begin{eqnarray}
\mu^2_{ren} &=&-2 (1-a) \label{bee1}\\ \lambda_{ren} &=& 4
(1-a)\label{bee2}
\end{eqnarray}
This yields the negative third-order Tschebyscheff polynomial
$-T_3(\Phi)= 3\Phi-4\Phi^3$. Equally well we can also obtain the
positive Tchebyscheff polynomial $T_3(\Phi )$ for the choice
\begin{eqnarray}
\mu^2_{ren} &=& 4(1-a) \\ \lambda_{ren} &=& -4(1-a)
\end{eqnarray}
The result is the chaotic $3B$ string ($N=3$, $b=0$, $s=1$).

We could also start from a $\phi^3$-theory, getting in a similar
way an $N=2$ string. Moreover, instead of doing the
nearest-neighbour coupling with $\Phi_n^{i-1}$ and $\Phi_n^{i+1}$,
we could equally well choose the updated variables
$T(\Phi_n^{i-1})$ and $T(\Phi_n^{i+1})$. This yields a chaotic
strings with $b=1$ rather than $b=0$. Finally, replacing $T$ by
$-T$ at odd lattice sites yields the anti-diffusive coupling form
$s=-1$.

\section{Appendix B: QCD formulas used}

In
a third-order QCD calculation the running strong coupling is given by
\newcommand{\lo}{{\ln \frac{E^2}{\Lambda^2}}}
\begin{eqnarray}
\alpha_s (E)&=&\frac{-2}{b_0 \lo}
\left\{ 1+\frac{2b_1 \ln \lo}{b_0^2 \lo }\right. \nonumber \\
&+& \left. \frac{4b_1^2}{b_0^4 \left( \lo \right)^2}
\left[ \left( \ln \lo -\frac{1}{2} \right)^2 +\frac{b_0b_2}{b_1^2}-\frac{5}{4}
\right] \right\} \nonumber \\
&+& O \left( \frac{1}{\left( \lo \right)^4 }\right)  . \label{runas}
\end{eqnarray}
Here $\Lambda$ is the QCD scale parameter, which takes on different values in the
various flavour regions, and the coefficients
$b_i$ are given by
\begin{eqnarray}
b_0 &=& -\frac{1}{2\pi} \left( 11 - \frac{2n_f}{3} \right) \\
b_1 &=& -\frac{1}{4\pi^2} \left( 51 - \frac{19n_f}{3} \right) \\
b_2 &=& -\frac{1}{64 \pi^3} \left( 2857 -\frac{5033n_f}{9}+
\frac{325 n_f^2}{27} \right)  .
\end{eqnarray}
$n_f$ denotes the relevant number of quark flavours. The integer
$n_f$ changes by 1 at thresholds given by the quark pole masses
$M_q$.
At the thresholds $M_q$, $\alpha_s (E)$ should be continuous. This
determines the scale parameters $\Lambda^{(n_f)}$ in the various
flavour regions, given one of the parameters $\Lambda^{(n_f)}$ for some
$n_f$ (or, equivalently, given $\alpha_s(E^*)$ at some fixed energy $E^*$).
The relation between pole quark masses $M_q$ and free quark masses $m_q$
is
\begin{equation}
M_q =m_q \left\{ 1+ \frac{4}{3} \frac{\alpha_s (M_q)}{\pi} +K_q
\left( \frac{\alpha_s (M_q)}{\pi} \right)^2 + O\left( \left(
\frac{\alpha_s (M_q)}{\pi} \right)^3 \right) \right\},
\label{pole}
\end{equation}
with
\begin{equation}
K_q=16.11-1.04 \sum_{i=1}^{n_f-1} \left( 1- \frac{M_i}{M_q}
\right)
\end{equation}
\cite{gray}. By free quark mass we actually mean the
$\overline{MS}$ running quark mass $\bar{m}_q(E)$ at
renormalization scale $E=M_q$. For the light quarks $u,d,s$ we
define the free masses to be the running $\overline{MS}$ masses at a
renormalization scale given by the proton energy scale $E\approx
1$ GeV.

\section*{Figure captions}

\pagestyle{empty}

\begin{description}

\item[Fig.~1] Feynman web interpretation of the chaotic string
dynamics.

\item[Fig.~2] Evolution of standard model coupling parameters to
stable zeros $a_1,a_2$ of the interaction energy.

\item[Fig.~3] Interaction energy $W(a)$ of the chaotic $3A$ string
in the region $a \in [0,1] $ (a) and $a \in [0,0.018]$ (b).
$W(a)$ is numerically obtained by iterating eq.~(\ref{dyn})
with $N=3$ and $b=1$
for random initial conditions and averaging 
$\frac{1}{2}\Phi_n^i\Phi_n^{i+1}$ over all $n$ and $i$.

\item[Fig.~4] Interaction energy $W(a)$
of the chaotic $3B$ string in the region $a \in [0,1]$ (a) and $a
\in [0,0.018]$ (b).

\item[Fig.~5] Interaction energy $W(a)$ of the chaotic $2A$ string.

\item[Fig.~6] Interaction energy $W(a)$ of the chaotic $2B$ string.

\item[Fig.~7] Interaction energy $W(a)$ of the chaotic $2A^-$
string.

\item[Fig.~8] Interaction energy $W(a)$ of the chaotic $2B^-$
string.

\item[Fig.~9] Self energy $V(a)$ of the chaotic $3A$ string in the region $a\in [0,1]$ (a)
and $a\in [0,0.022]$ (b). $V(a)$ is numerically obtained
by iterating eq.~(\ref{dyn}) with $N=3$ and $b=1$
for random initial conditions
and averaging $\frac{3}{2}(\Phi_n^i)^2-(\Phi_n^i)^4$ over all $n$
and $i$.

\item[Fig.~10] Self energy $V(a)$ of the chaotic $3B$ string.

\item[Fig.~11] Self energy $V(a)$ of the chaotic $2A$ string.
$V(a)$ is obtained by iterating eq.~(\ref{dyn}) with $N=2$ and $b=1$
and averaging 
$\Phi_n^i-\frac{2}{3}(\Phi_n^i)^3$ over all $n$ and $i$.

\item[Fig.~12] Self energy $V(a)$ of the chaotic $2B$ string.

\item[Fig.~13] $|\log |V(a)-V(0)||$ versus $\log a$ for the $3A/B$ string in the scaling region.

\item[Fig.~14] $|\log |V(a)-V(0)||$ versus $\log a$ for the $2A/B$ string in the scaling region.

\item[Fig.~15] One period of the self energy of the $2A/2B$
string in the scaling region. The
various minima fix the fermion masses.

\item[Fig.~16] Same as Fig.~15, but for the $2A^-/2B^-$ string.
The minima fix the boson masses.

\item[Fig.~17] Summary of the way in which the 2 types of vacuum energies of
the 6 types of chaotic string theories fix standard model parameters,
(a) interaction energy $W(a)$, (b) self energy $V(a)$.

\item[Fig.~18] A possible correspondence between superstrings
and chaotic strings. The left hand side shows the 5 known
superstring theories plus the 11-dimensional theory, which
compactified on a circle is dual to the IIA string and
compactified on an interval dual to the $HET\; E8\times E8$
string. The right hand side shows a similar diagram for the
chaotic string theories.

\end{description}

\newpage
\epsfig{file=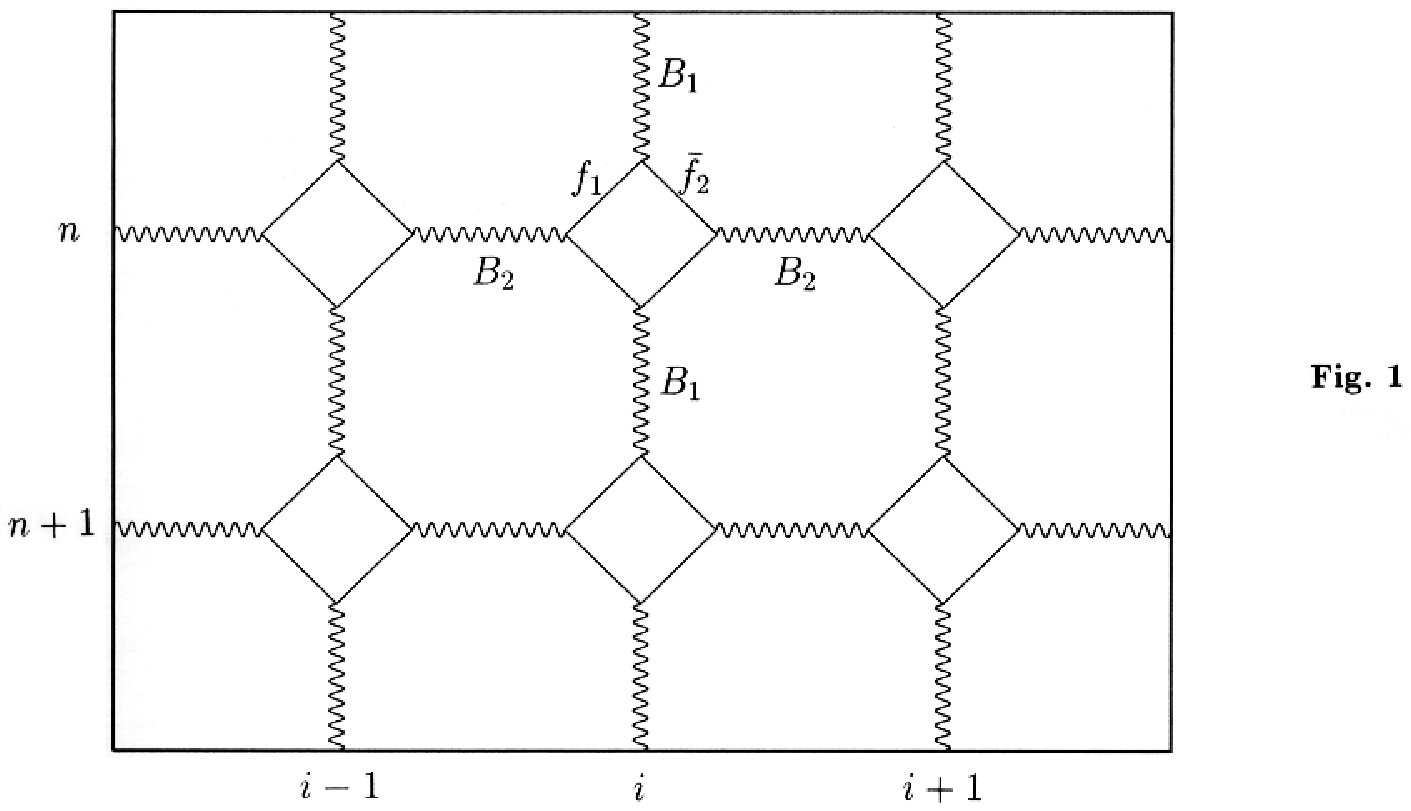}
\newpage
\epsfig{file=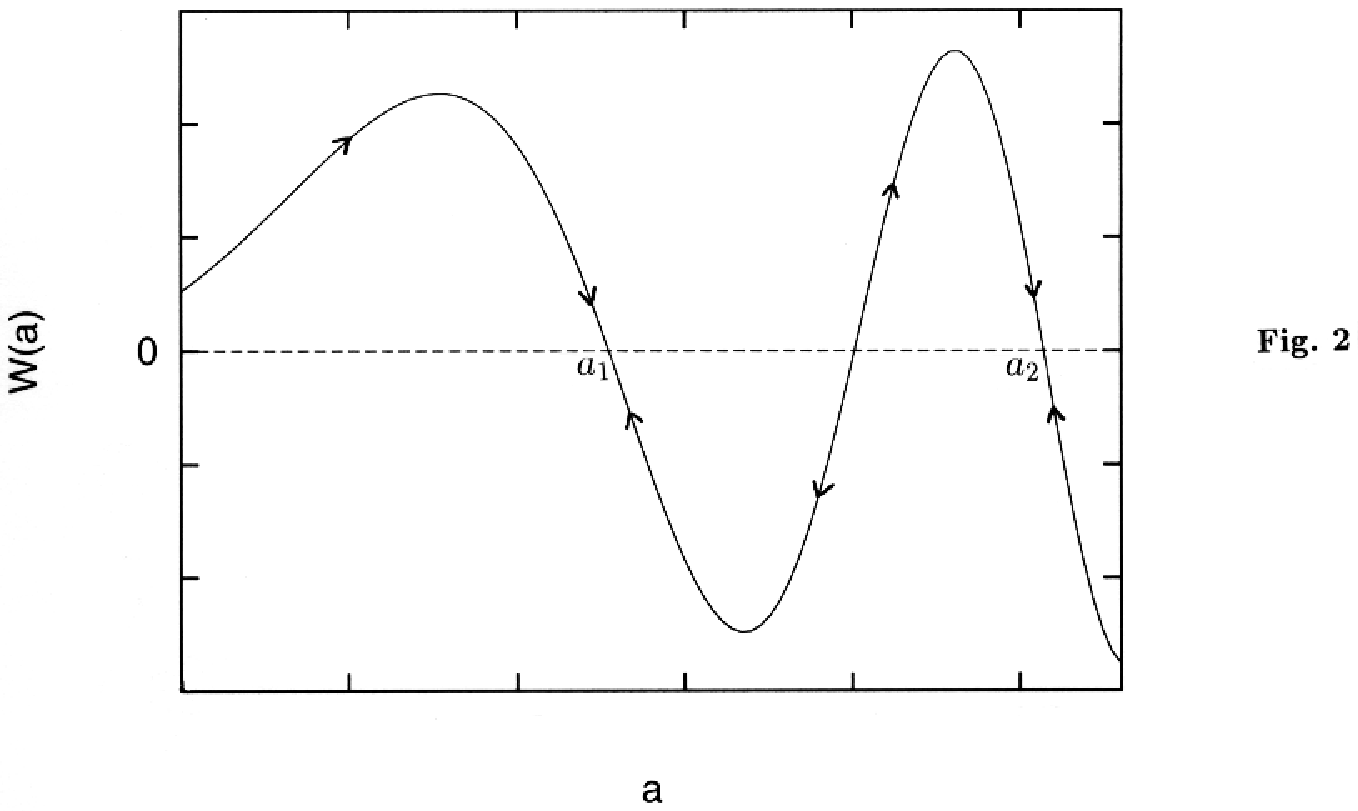}
\newpage
\epsfig{file=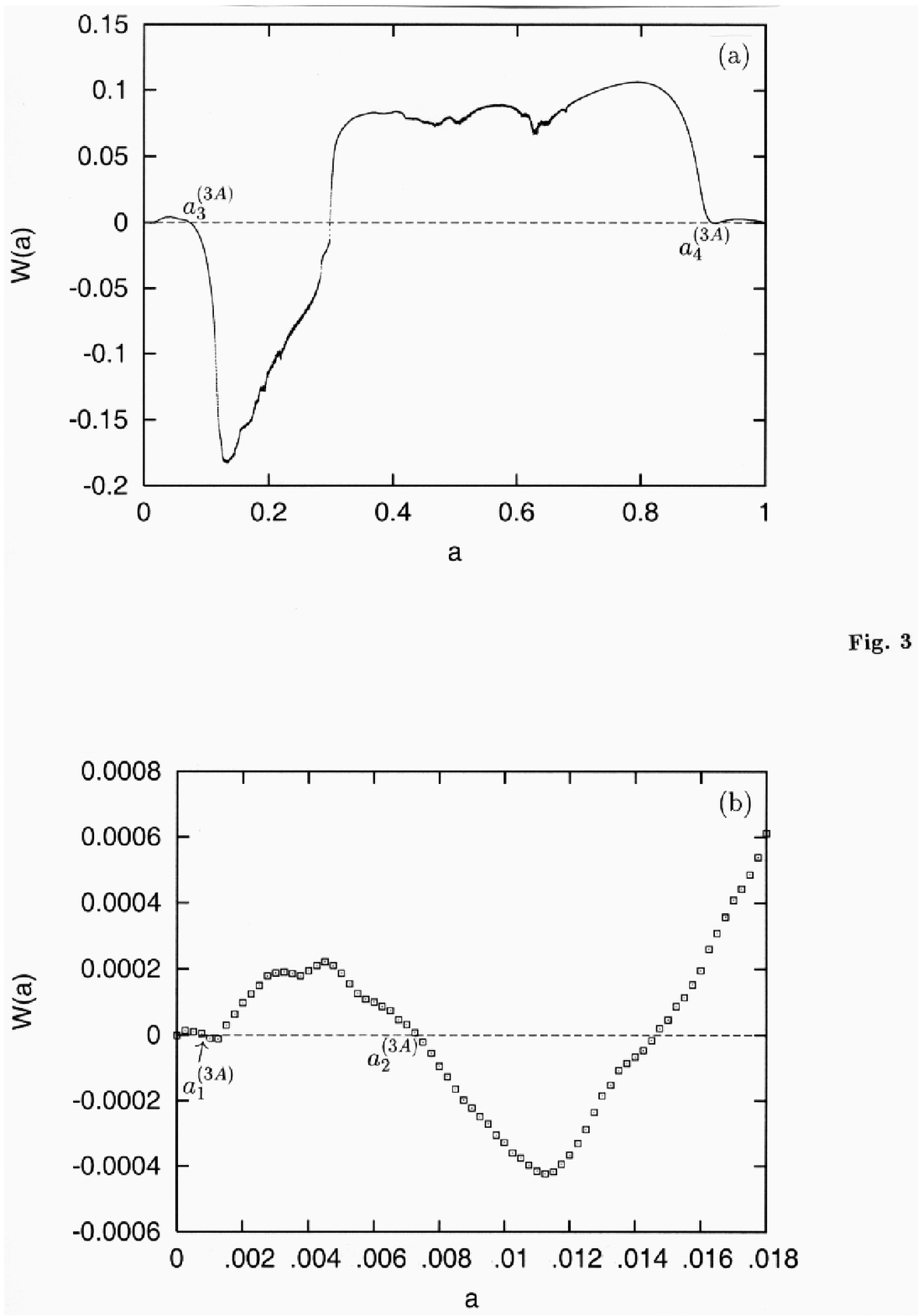}
\newpage
\epsfig{file=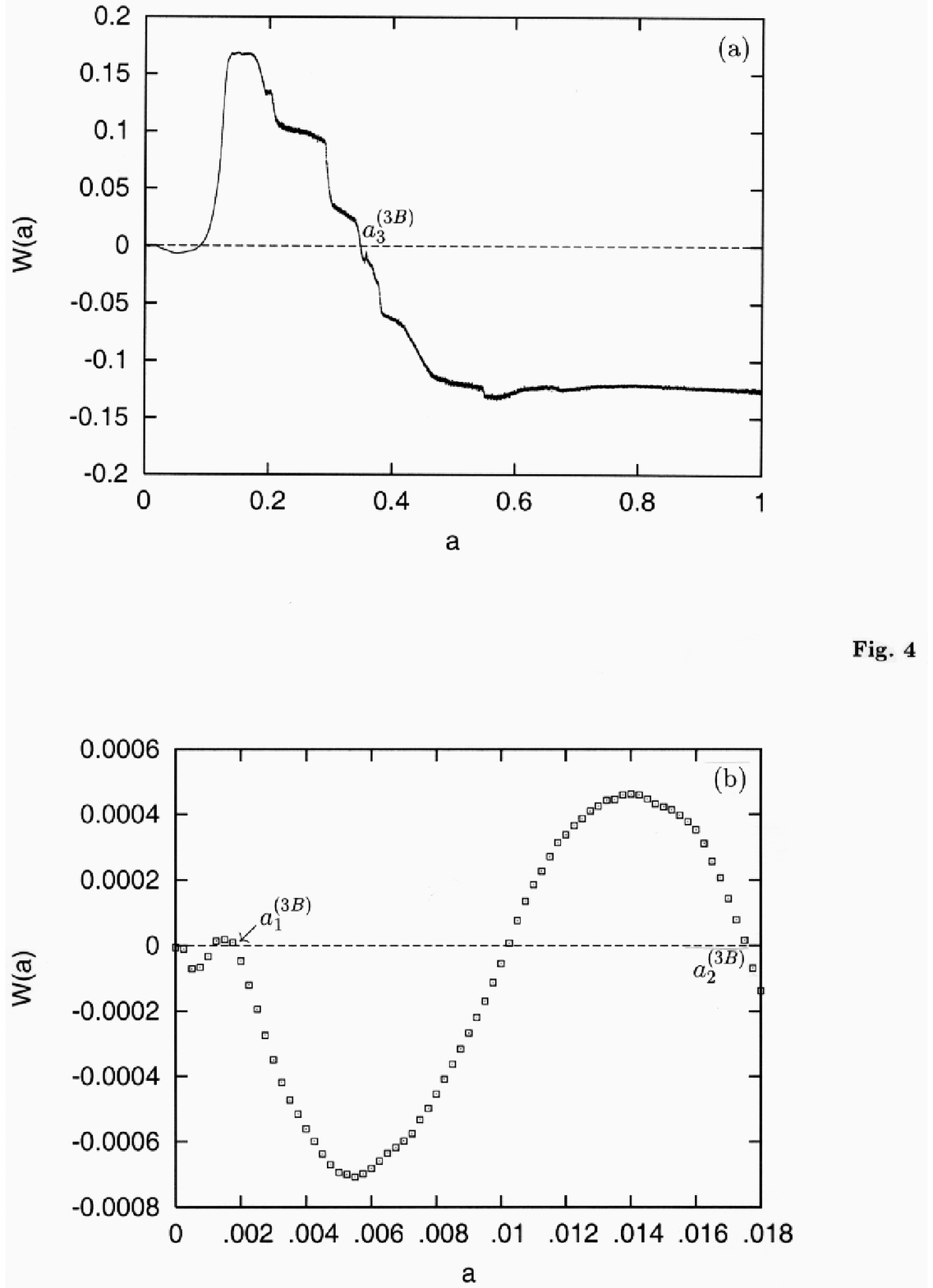}
\newpage
\epsfig{file=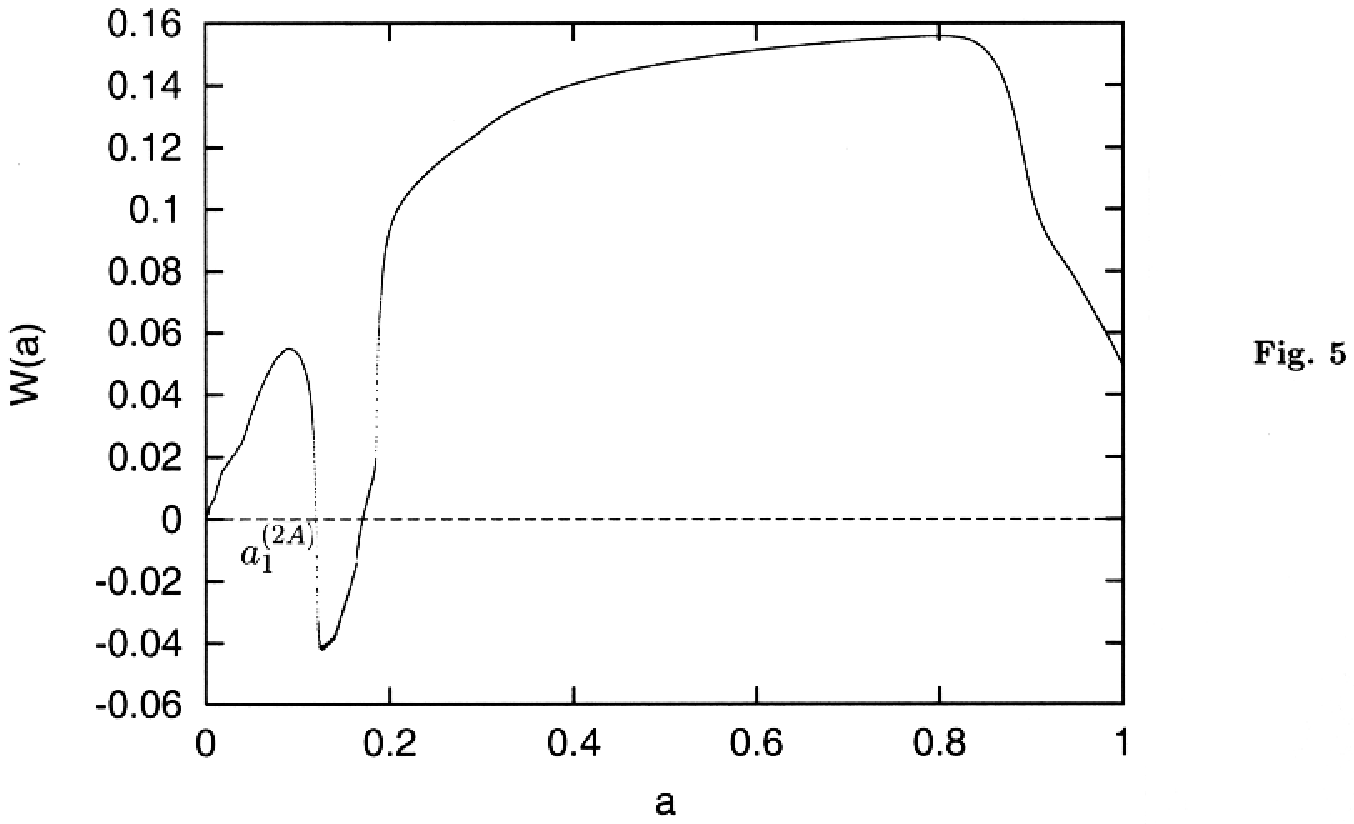}
\newpage
\epsfig{file=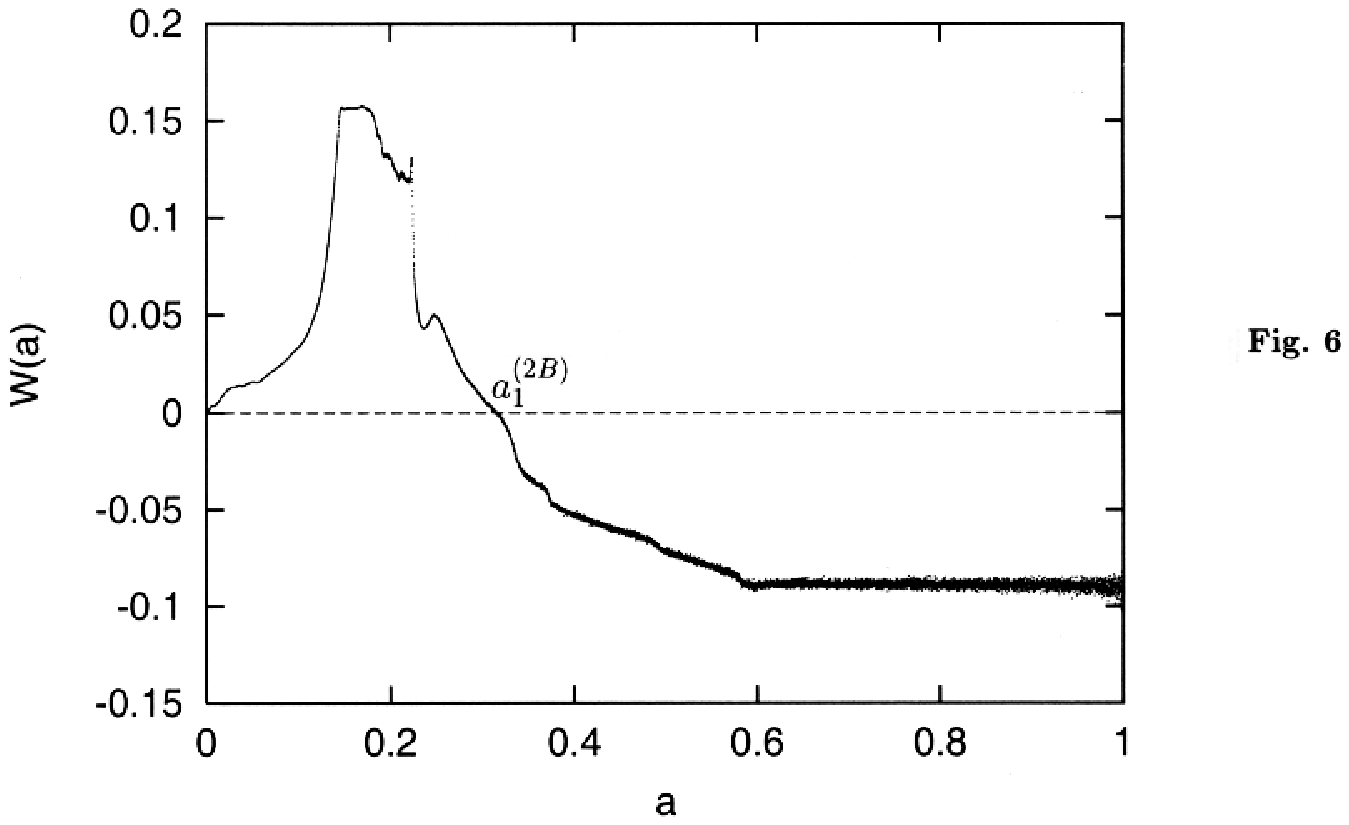}
\newpage
\epsfig{file=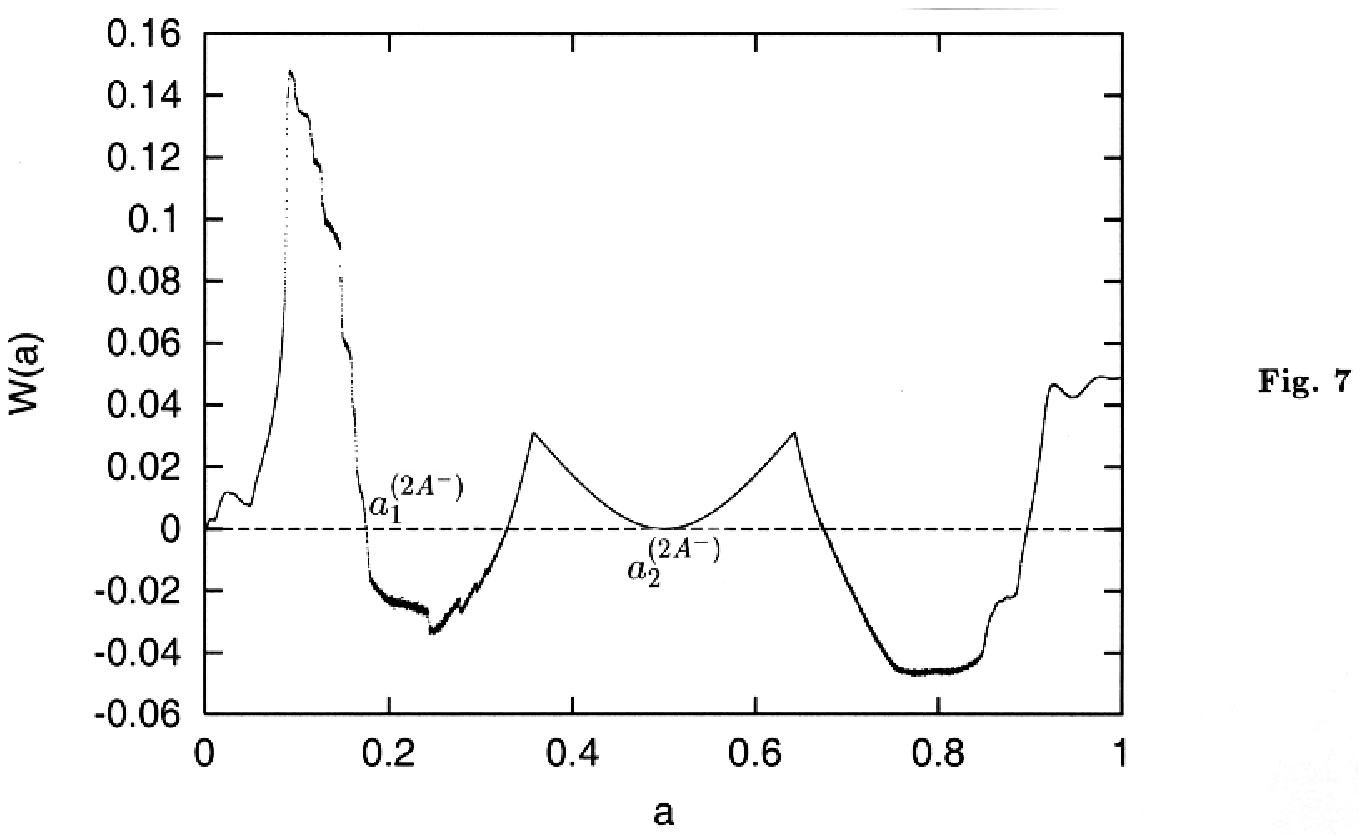}
\newpage
\epsfig{file=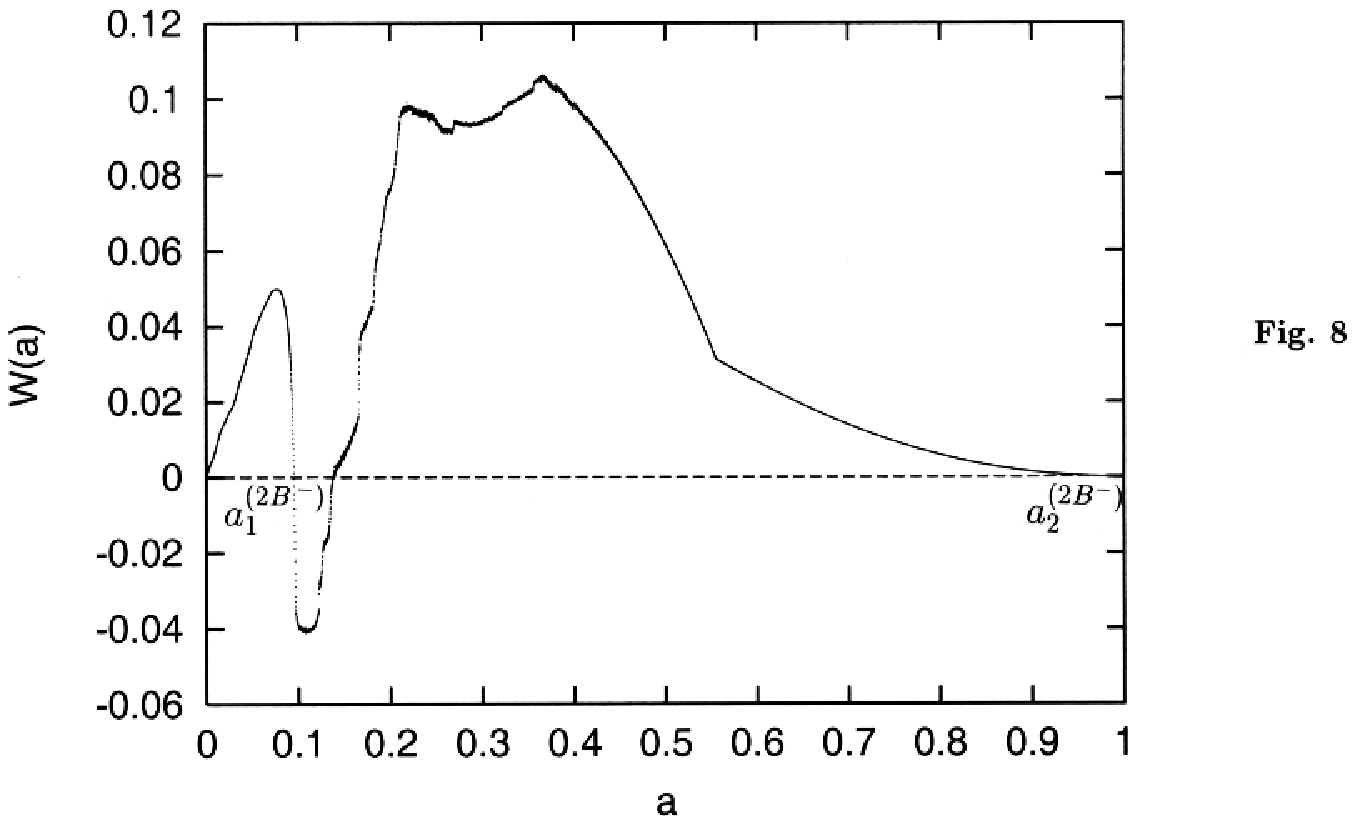}
\newpage
\epsfig{file=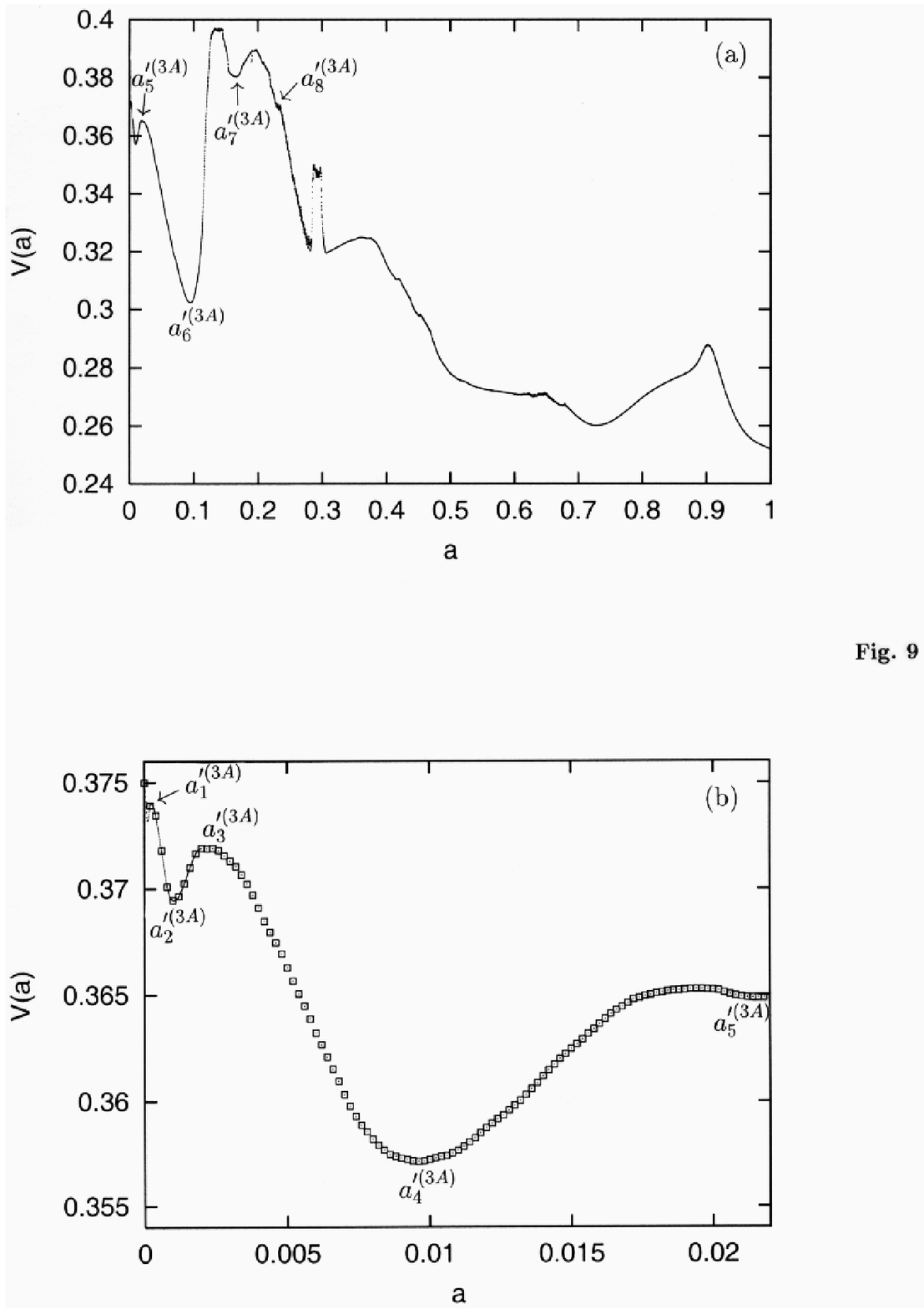}
\newpage
\epsfig{file=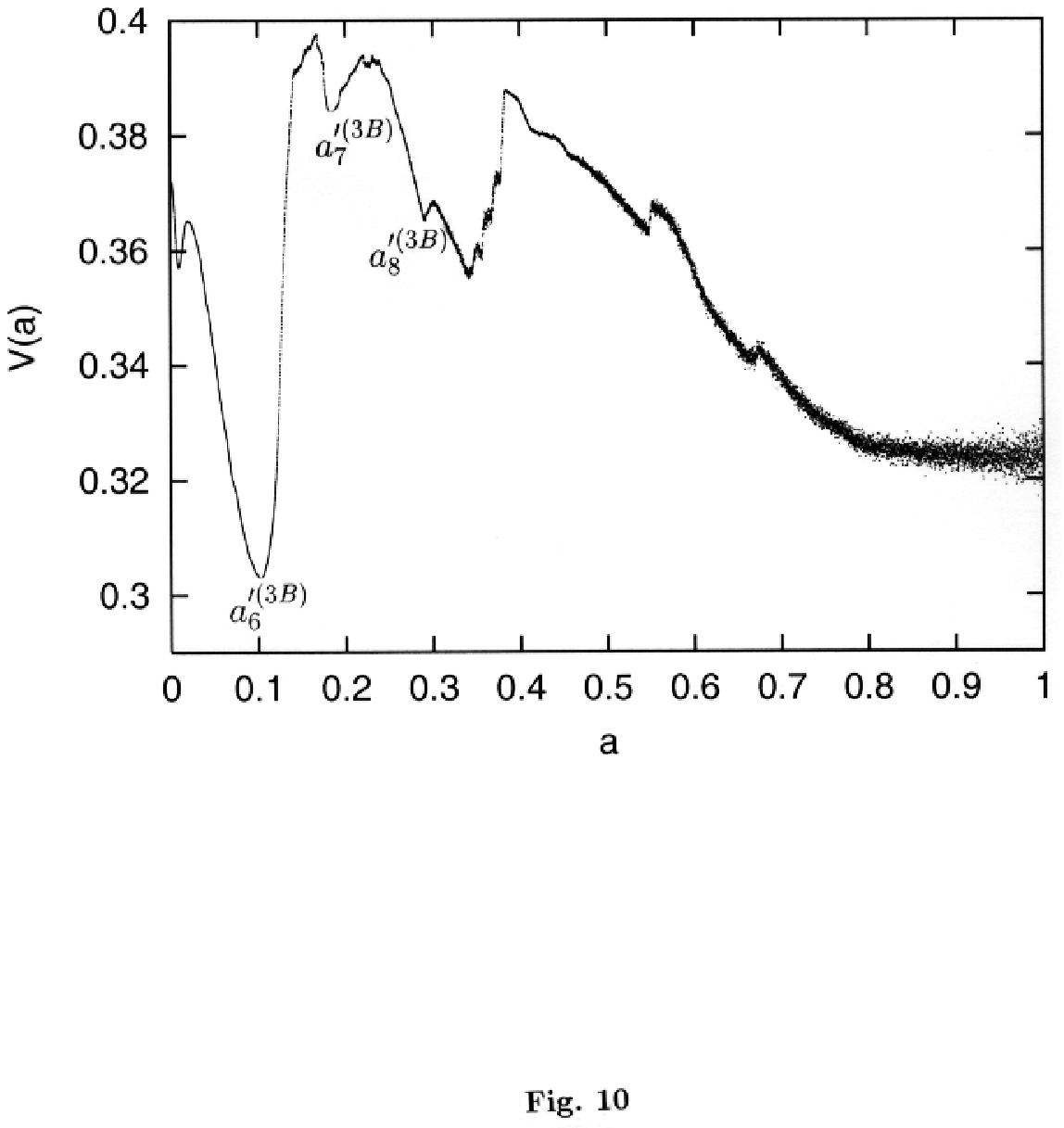}
\newpage
\epsfig{file=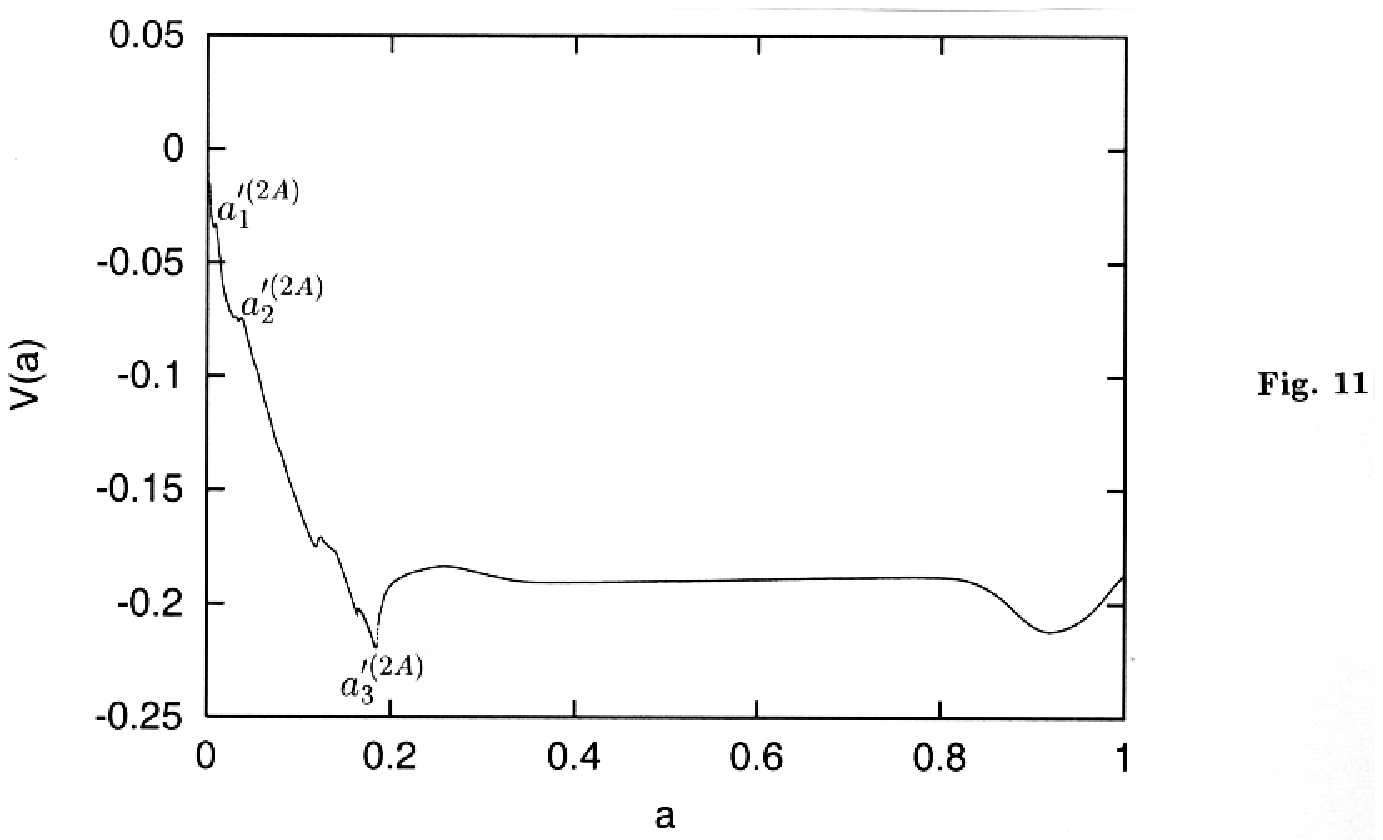}
\newpage
\epsfig{file=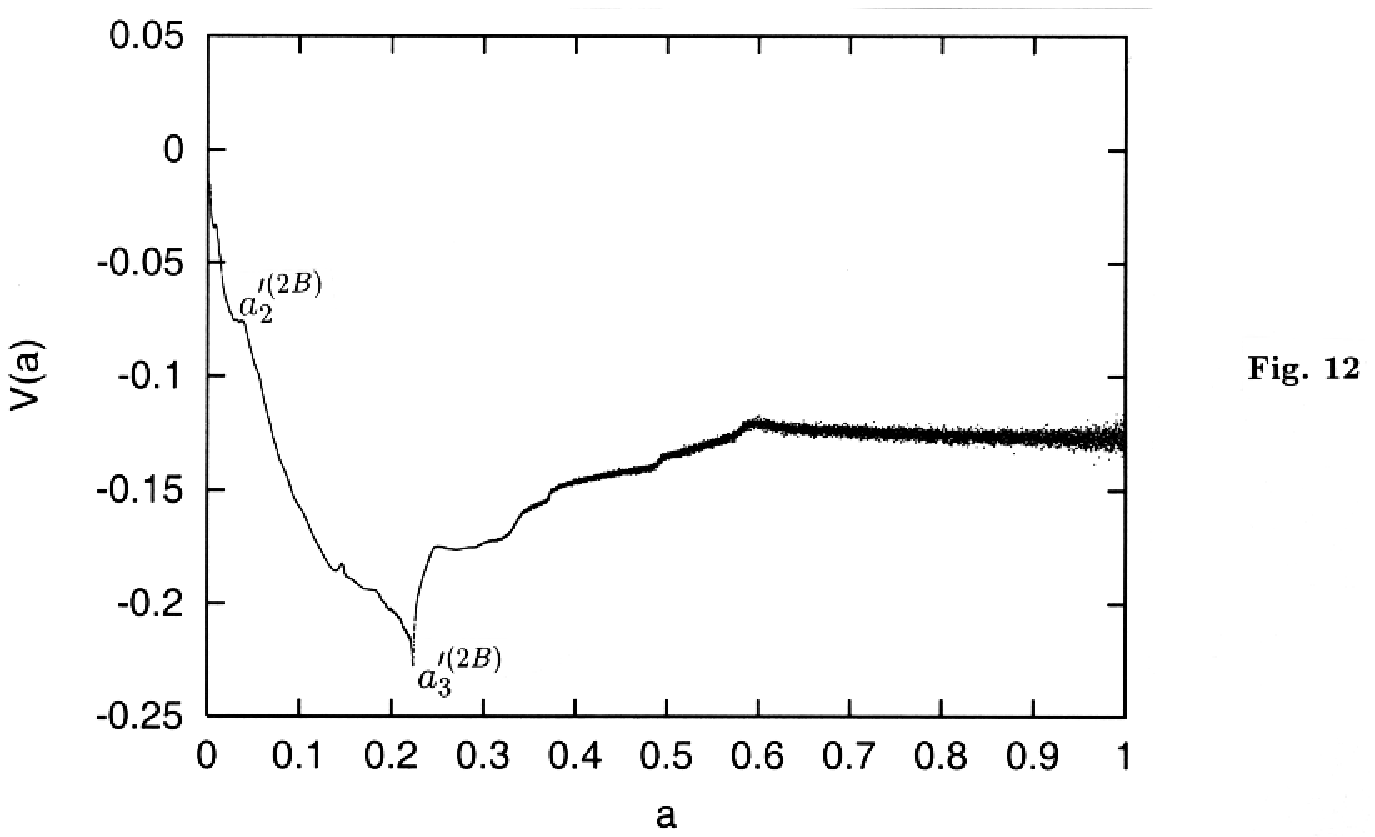}
\newpage
\epsfig{file=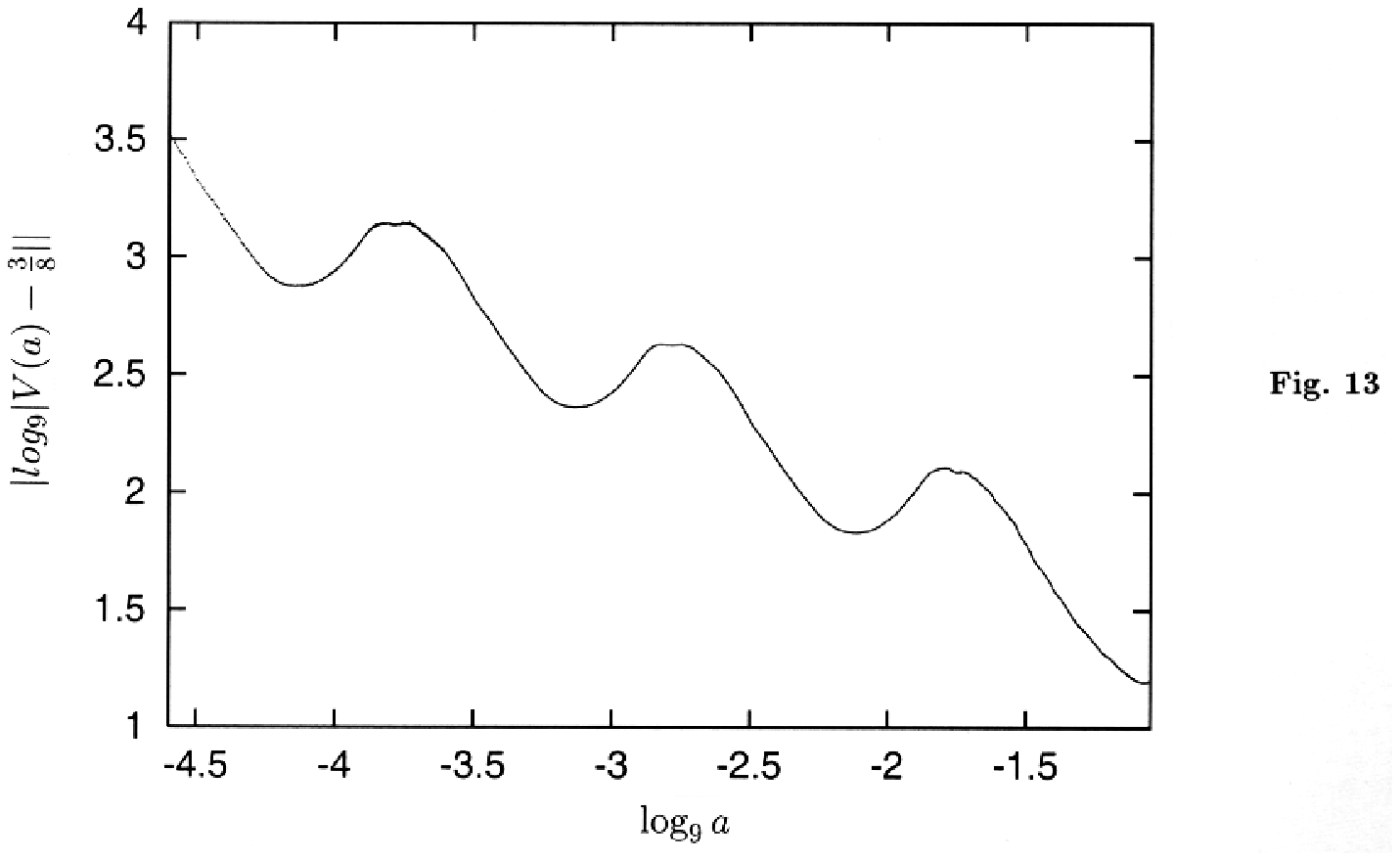}
\newpage
\epsfig{file=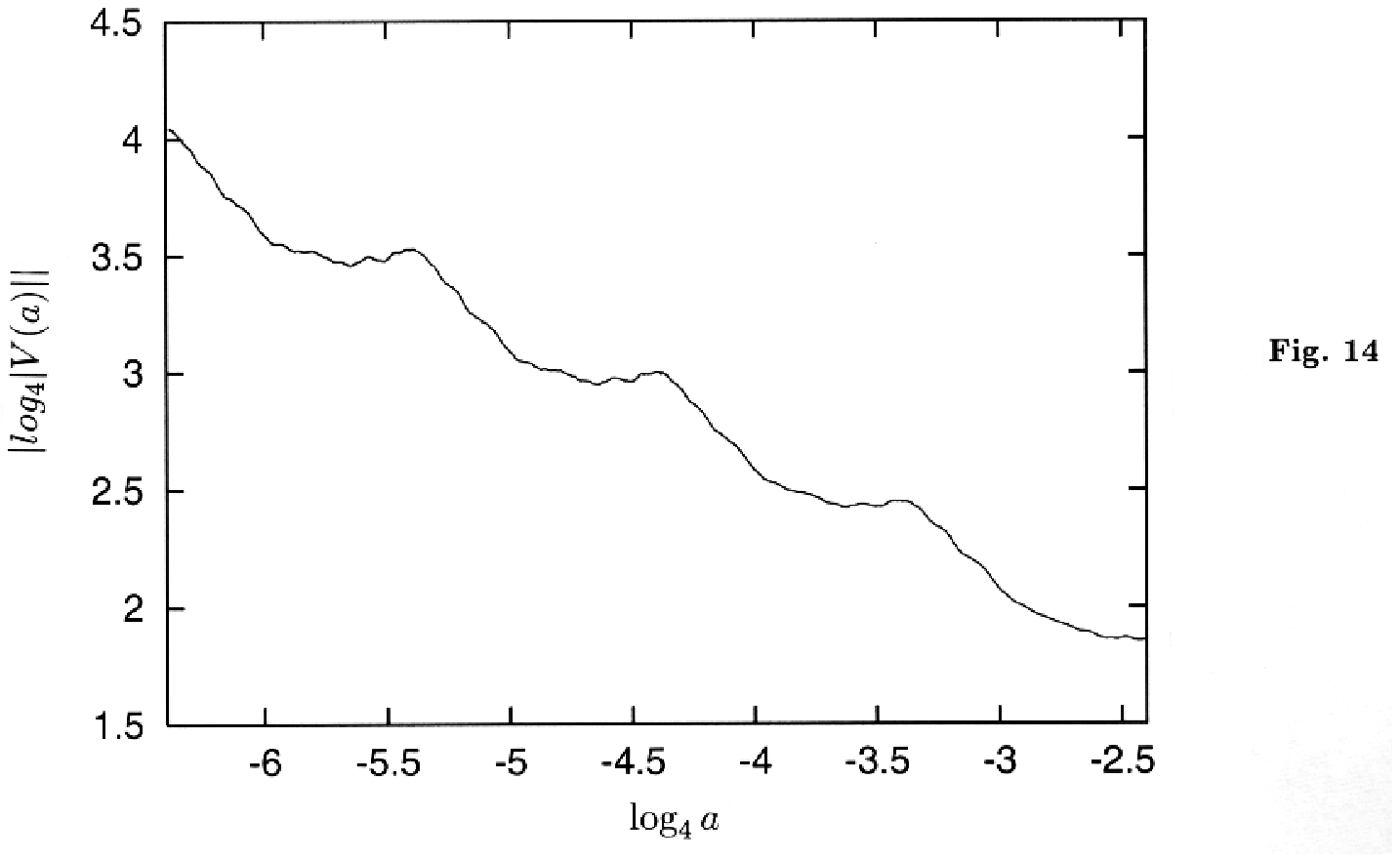}
\newpage
\epsfig{file=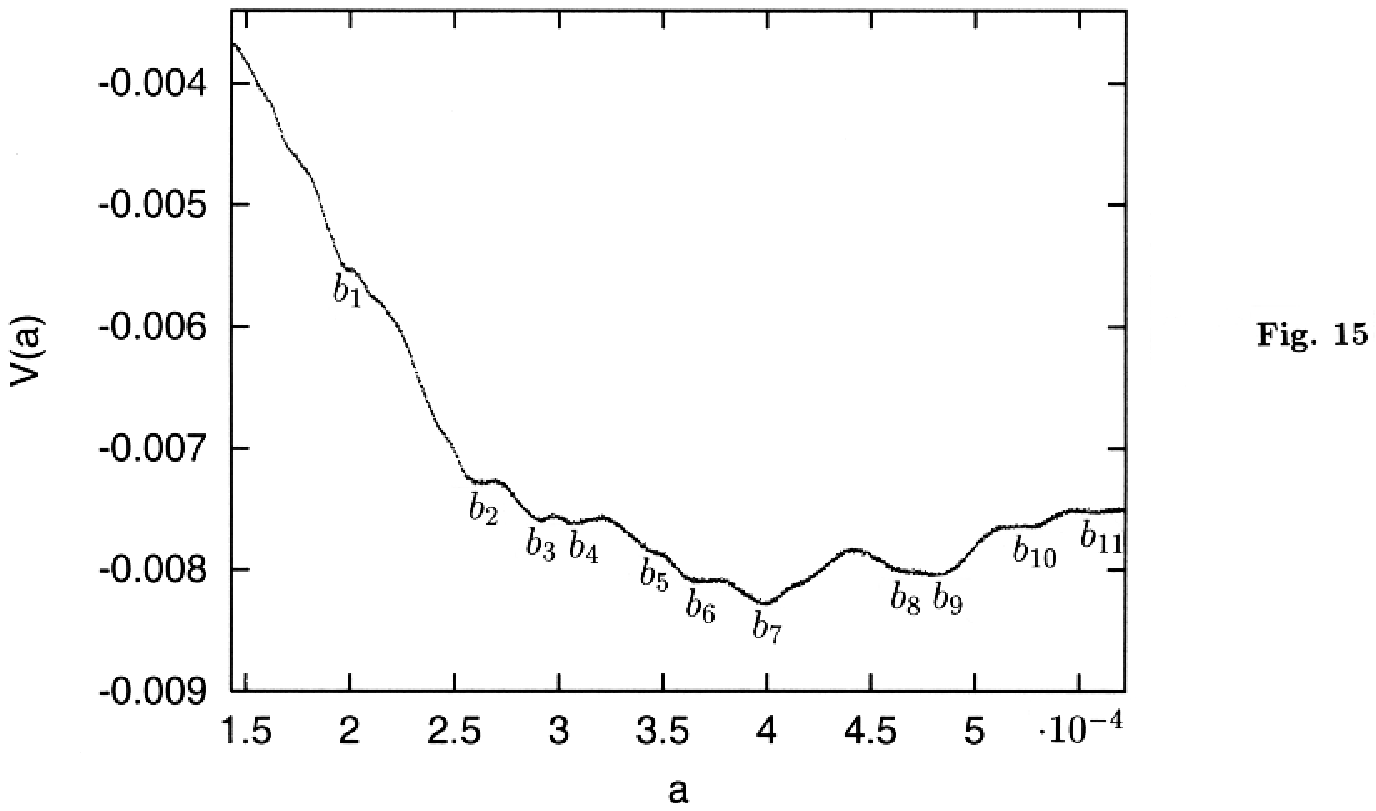}
\newpage
\epsfig{file=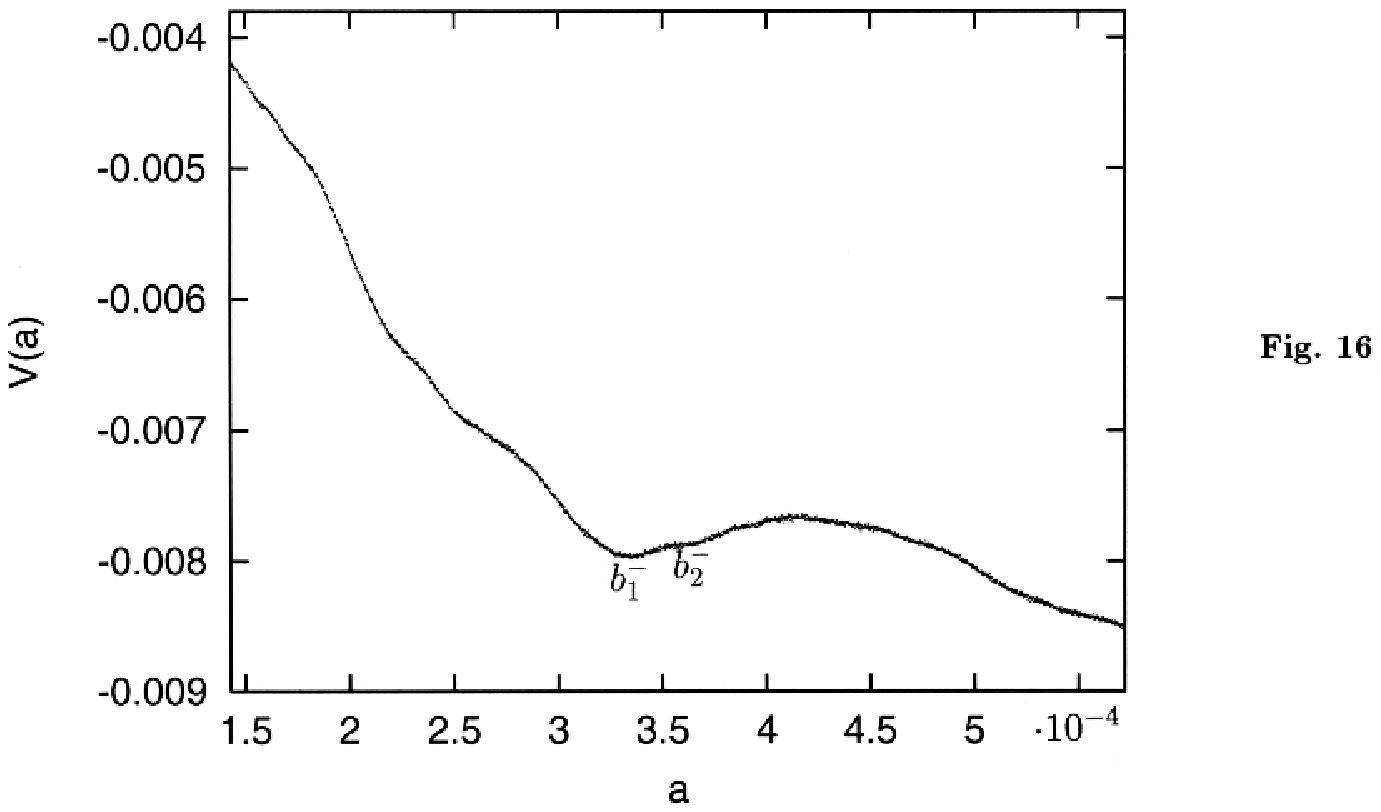}
\newpage
\epsfig{file=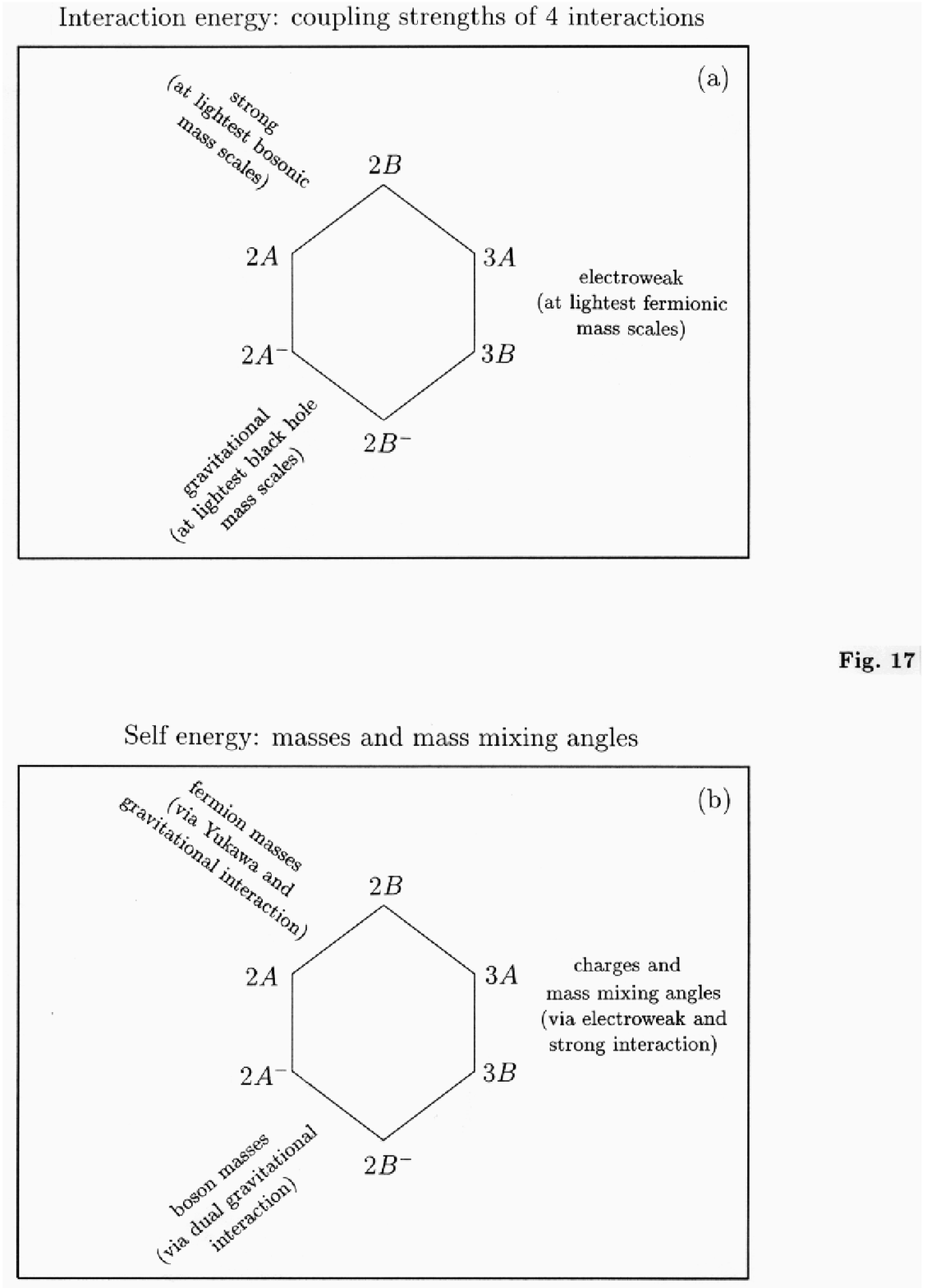}
\newpage
\epsfig{file=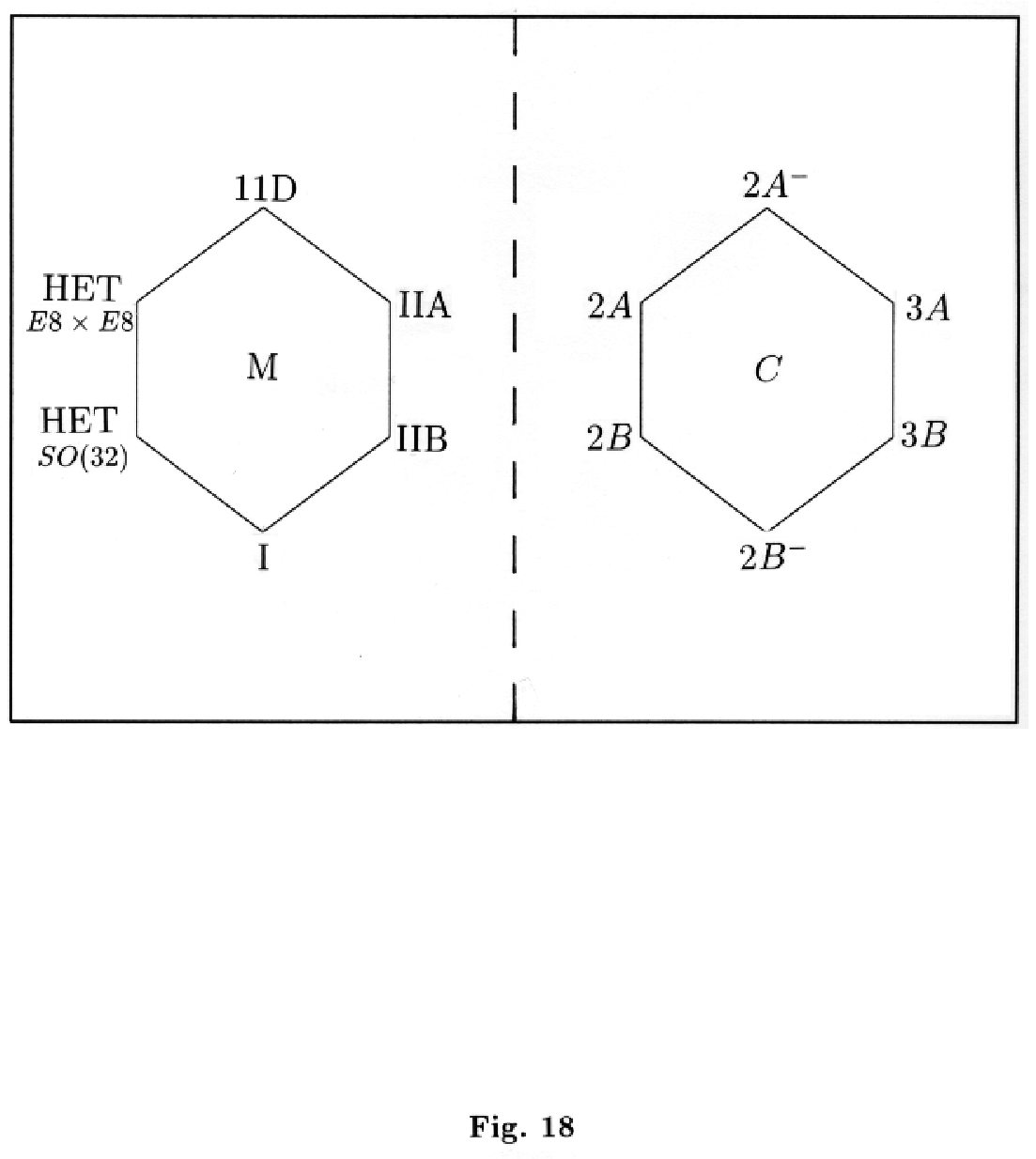}

\end{document}